\documentclass[pra,nofootinbib,floats,aps,superscriptaddress,twocolumn]{revtex4}
\pdfoutput=1 
\usepackage{graphicx}
\usepackage{amssymb}
\usepackage{amsmath}
\usepackage{amsfonts}
\usepackage[usenames]{color}
\usepackage{amsbsy}
\usepackage{epsfig}
\newcommand{\cc}{{\mbox{c.c.\,}}}
\newcommand{\cZ}{\mathcal Z}
\newcommand{\cU}{\mathcal U}

\def\tr{\,{\rm tr}\,}

\newcommand{\be}{\begin{equation}}
\newcommand{\ee}{\end{equation}}
\newcommand{\bea}{\begin{eqnarray}}
\newcommand{\eea}{\end{eqnarray}}
\newcommand{\ben}{\begin{enumerate}}
\newcommand{\een}{\end{enumerate}}
\newcommand{\bit}{\begin{itemize}}
\newcommand{\eit}{\end{itemize}}

\numberwithin{equation}{section}

\def\a{\alpha}
\def\b{\beta}

\newcommand{\cE}{\ensuremath{{\cal E}}}
\definecolor{BrickRed}{cmyk}{0,0.89,0.94,0.28}
\definecolor{MidnightBlue}{cmyk}{0.98,0.13,0,0.43}
\definecolor{DarkGreen}{rgb}{0,0.7,0.1}

\newcommand{\bfx}{{\bf x}} 
\newcommand{\bfX}{{\bf X}} 
\newcommand{\bfy}{{\bf y}}
\newcommand{\bfz}{{\bf z}}

\newcommand{\bA}{\mathbf A}
\newcommand{\bE}{\mathbf E}
\newcommand{\bB}{\mathbf B}
\newcommand{\bD}{\mathbf D}
\newcommand{\bJ}{\mathbf J}
\newcommand{\bH}{\mathbf H}
\newcommand{\bM}{\mathbf M}
\newcommand{\bN}{\mathbf N}

\newcommand{\bI}{\mathbf I}
\newcommand{\bQ}{\mathbf Q}
\newcommand{\bP}{\mathbf P}
\newcommand{\bphi}{\mbox{\boldmath$\phi$}}

\newcommand{\cC}{{\mathcal C}}
\newcommand{\cF}{{\mathcal F}}

\newcommand{\cD}{{\mathcal D}}
\newcommand{\cT}{{\mathcal T}}
\newcommand{\cG}{{\mathcal G}}

\begin{document}

\title{Fluctuation induced quantum interactions between compact objects\\ and a plane mirror}
 
\author{T.\ Emig}
\affiliation{Institut f\"ur Theoretische Physik, Universit\"at zu K\"oln,
Z\"ulpicher Strasse 77, 50937 K\"oln, Germany}
\affiliation{Laboratoire de Physique Th\'eorique et Mod\`eles
Statistiques, CNRS UMR 8626, Universit\'e Paris-Sud, 91405 Orsay,
France}

\begin{abstract} 
  The interaction of compact objects with an infinitely extended
  mirror plane due to quantum fluctuations of a scalar or
  electromagnetic field that scatters off the objects is studied. The
  mirror plane is assumed to obey either Dirichlet or Neumann boundary
  conditions or to be perfectly reflecting. Using the method of
  images, we generalize a recently developed approach for compact
  objects in unbounded space \cite{Emig+07b,Emig+07} to show that the
  Casimir interaction between the objects and the mirror plane can be
  accurately obtained over a wide range of separations in terms of
  charge and current fluctuations of the objects and their images. Our
  general result for the interaction depends only on the scattering
  matrices of the compact objects. It applies to scalar fields with
  arbitrary boundary conditions and to the electromagnetic field
  coupled to dielectric objects. For the experimentally important
  electromagnetic Casimir interaction between a perfectly conducting
  sphere and a plane mirror we present the first results that apply at
  all separations. We obtain both an asymptotic large distance
  expansion and the two lowest order correction terms to the proximity
  force approximation. The asymptotic Casimir-Polder potential for an
  atom and a mirror is generalized to describe the interaction between
  a dielectric sphere and a mirror, involving higher order multipole
  polarizabilities that are important at sub-asymptotic distances.
\end{abstract}

\maketitle

\section{Introduction}

The interaction between neutral objects is dominated by fluctuation
forces due to the coordinated behavior of fluctuating charges or
collective modes inside the objects.  At zero temperature or at
sufficiently small distances, the interactions result from quantum
fluctuations. They are important on atomic scales as well as between
macroscopic bodies. A prominent example for the latter is the Casimir
force between parallel metallic plates due to current or
electromagnetic field fluctuations \cite{Casimir:1948dh}. During the
last decade experimental verifications of this effect have been
performed with increasing precision. These high precision measurements
were enabled by the use of curved surfaces instead of parallel planar
mirrors in order to avoid the problem of parallelism. The most
commonly employed geometry is a sphere-plate setup that was used in
the first high-precision tests of the Casimir effect
\cite{Lamoreaux+97,Mohideen+98}. This geometry has been successfully
used ever since in most of the recent experimental studies of Casimir
forces between metallic surfaces
\cite{Roy+99,Ederth00,Chan+01,Chen+02,Chen:2006c,Decca:2007a,Chen:2007b,Munday:2007a}.
In order to keep the deviations from two parallel plates sufficiently
small, spheres with a radius much larger than the surface distance
have been used.  The effect of curvature has been accounted for by the
``proximity force approximation'' (PFA) \cite{Parsegian}. This scheme
is assumed to describe the interaction for sufficiently small ratios
of radius of curvature to distance. However, this an uncontrolled
assumption since PFA becomes exact only for infinitesimal separations,
and corrections to PFA are generally unknown.

At the other extreme, the interaction between a planar surface and
objects that are either very small or at asymptotically large distance
is governed by the Casimir-Polder potential that was derived for the
case of an atom and a perfectly conducting plane
\cite{Casimir+48}. This limit has been probed
experimentally with high precision for a Bose-Einstein condensate that
was trapped close to a planar surface \cite{Harber:2005}. Recently,
there have been first attempts to go beyond the two extreme limits of
asymptotically large and small separations by measuring the Casimir
force between a sphere and a plane over a larger range of ratios of
sphere radius to distance \cite{Krause:2007a}.

So far, no theoretical prediction is available that can describe
the electromagnetic Casimir interaction between a compact object and a
planar surface at all distances, including the important sphere-plate
geometry. Until recently, progress in understanding the geometry
dependence of fluctuation forces was hampered by the lack of practical
methods that are applicable at all separations. Conceptually, the
effect of geometry and shape is difficult to study due to the
non-additivity of fluctuation forces. Explicit consequences of this
non-additivity and also non-monotonic changes in the force have
been recently predicted for a pair of cylinders next to
planar walls \cite{Rodriguez:2007a+b}.  This behavior has been
interpreted in terms of collective charge fluctuations inside the
bodies \cite{Rahi:2007a}.

For some decades, there has been considerable interest in the theory
of Casimir forces between objects with curved surfaces. Two types of
approaches have been pursued. Attempts to compute the force explicitly
in particular geometries and efforts to develop a general framework
which yields the interaction in terms of characteristics of
the objects like polarizability or curvature. Within the second type
of approach, Balian and Duplantier studied the electromagnetic
Casimir interaction between compact perfect metals in terms of a
multiple reflection expansion and derived also explicit results to
leading order at asymptotically large separations \cite{Balian}.  For
parallel and partially transmitting plates a connection to
scattering theory has been established which yields the Casimir
interaction of the plates as a determinant of a diagonal matrix of
reflection amplitudes \cite{Jaeckel:1991}. For non-planar, deformed
plates, a general representation of the Casimir energy as a functional
determinant of a matrix that describes reflections at the surfaces and
free propagation between them has been developed in
Ref.~\cite{Emig+04}. Later on, an equivalent representation has been
applied to perturbative computations in the case of rough and corrugated
plates with finite conductivity \cite{Neto+05,Lambrecht+06}. 

Functional determinant formulas have been used also for open
geometries that do not fall into the class of parallel plates with
deformations.  For the electromagnetic Casimir interaction between a
planar plate and infinitely long cylinders, a partial wave expansion
of the functional determinant has been developed \cite{Emig+06}.  The
same results have been reobtained and used to compute corrections to
the PFA for the cylinder-plate geometry in Ref.~\cite{Bordag06}.
Kenneth and Klich identified the inverted Green's function in the
functional determinant as a T-matrix and derived a formal result for
the Casimir interaction in terms of this matrix in the case of scalar
fields in a medium with a space and frequency dependent speed of light
\cite{Kenneth+06}.

Recently we described a new method based on a multipole expansion of
fluctuating charges that makes possible accurate and efficient
calculations of Casimir forces and torques between any number of
compact objects \cite{Emig+07b,Emig+07}.  The method applies to
electromagnetic fields and dielectrics as well as perfect conductors.
It also applies to other fields, such as scalar and Dirac, and to any
boundary conditions.  In this approach, the Casimir energy is given in
terms of the fluctuating field's scattering amplitudes from the
individual objects, which encode the effects of the shape and boundary
conditions. An equivalent partial wave expansion has been applied to a
scalar model for dielectrics \cite{Kenneth+07}. 

Casimir interactions due to scalar field fluctuations serve as a
simplified model for the full electromagnetic interaction. This model
is usually easier to analyze and provides an important tool in
developing conceptually new approaches and in estimating geometry
dependencies. For the sphere-plate geometry with Dirichlet boundary
conditions Bulgac {\it et al.} obtained the Casimir interaction over a
wide range of separations from a modified Krein trace formula
\cite{Bulgac+06,Wirzba:qfext07}. Most notably in the context of scalar
fields, a versatile numerical world-line algorithm based on Monte
Carlo methods has been developed and applied to a number of
interesting geometries, including the here studied sphere-plate
interaction \cite{Gies+06a,Gies+06b}.

In this work, we extend our approach developed in
Refs.~\cite{Emig+07b,Emig+07} to describe the interaction of compact
objects in the presence of a plane mirror. Our general result holds
for scalar and electromagnetic fields. In the electromagnetic case the
mirror is assumed to be perfectly conducting but the compact objects
can be dielectrics or perfect conductors. The derivation of the
general result for the Casimir interaction, see
Eqs.~\eqref{eq:Casimir-energy-gen-result} and
\eqref{eq:EM-Casimir-energy-gen-result}, is given in Section
\ref{sec:derivatiobn} by combining a functional integral approach and
the method of images. In Section \ref{sec:plate-sphere} we apply
our approach to compute the interaction of a sphere with a plane mirror
over a wide range of separations for the scalar Dirichlet and Neumann problem
and for the electromagnetic field that is most relevant to applications.
We provide a large distance expansion of the interaction, generalizing the
Casimir-Polder potential to include higher order multipole polarizabilities.
For small separations, we compute the two leading correction terms to the PFA.

\section{Interaction between fluctuating sources and their images}
\label{sec:derivatiobn}

In this Section we first review the functional integral formulation of
Casimir interactions between compact objects for a scalar field and
the electromagnetic field.  The fluctuating field is integrated out in
order to obtain an effective action for the fluctuating sources on the
objects. Then we show that the interaction of compact objects with an
infinite plane mirror can be described by the equivalent problem of
the interaction of compact objects with their mirror images in
otherwise empty space, i.e., without the plane mirror. This
equivalence holds for Dirichlet or Neumann boundary conditions (at the
mirror and the objects) for a scalar field and for a perfectly
conducting mirror plane and arbitrary dielectric objects in the case
of electromagnetic fluctuations.

\subsection{Scalar field}

We first consider a real quantum field $\phi(\bfx,t)$ in order to
introduce the method of images in the path integral formulation of
Casimir interactions. We assume that the space is divided into two
half spaces by a mirror plane at $z=0$. $N$ fixed closed surfaces
$\Sigma_\alpha$, $\alpha=1,\ldots,N$, are located in the right half
space ($z>0$). These surfaces are regarded as the boundaries of objects
on which either Dirichlet or Neumann boundary conditions are imposed.
The action of the unconstrained field in Minkowski space is
\begin{equation}
  \label{eq:action-minkowski}
  S[\phi] = \frac{1}{2}\int dt \int_> d\bfx \left\{ \frac{1}{c^2} (\partial_t \phi)^2 
- (\nabla\phi)^2\right\} \, .
\end{equation}
Here the $\bfx$-integration runs over the right half space, indicated
by $\int_>$.  The free energy $\cF$ of the constraint field
$\phi(\bfx,t)$ at inverse temperature $\beta$ is represented by the
Euclidean functional integral
\begin{equation}
  \label{eq:free-energy-func-int}
  e^{-\beta\cF} = Z = \int [\cD\phi]_\cC \exp\left( -S_E[\phi]/\hbar \right) \, ,
\end{equation}
where the Euclidean action $S_E$ follows from
Eq.~\eqref{eq:action-minkowski} after the Wick rotation $ct = -i x^0$,
\begin{equation}
  \label{eq:action-euclidean}
  S_E = \frac{1}{2c} \int_0^{\Lambda} dx^0 \int_> d\bfx \left\{
(\partial_0 \phi)^2 + (\nabla\phi)^2 \right\} 
\end{equation}
with $\Lambda=\beta\hbar c$.  The functional integral is over all
fields that are periodic in the $x_0$-interval from $0$ to $\Lambda$
and that obey the boundary conditions at the surfaces, indicated by
the subscript $\cC$.

\begin{figure}
\includegraphics[scale=0.8]{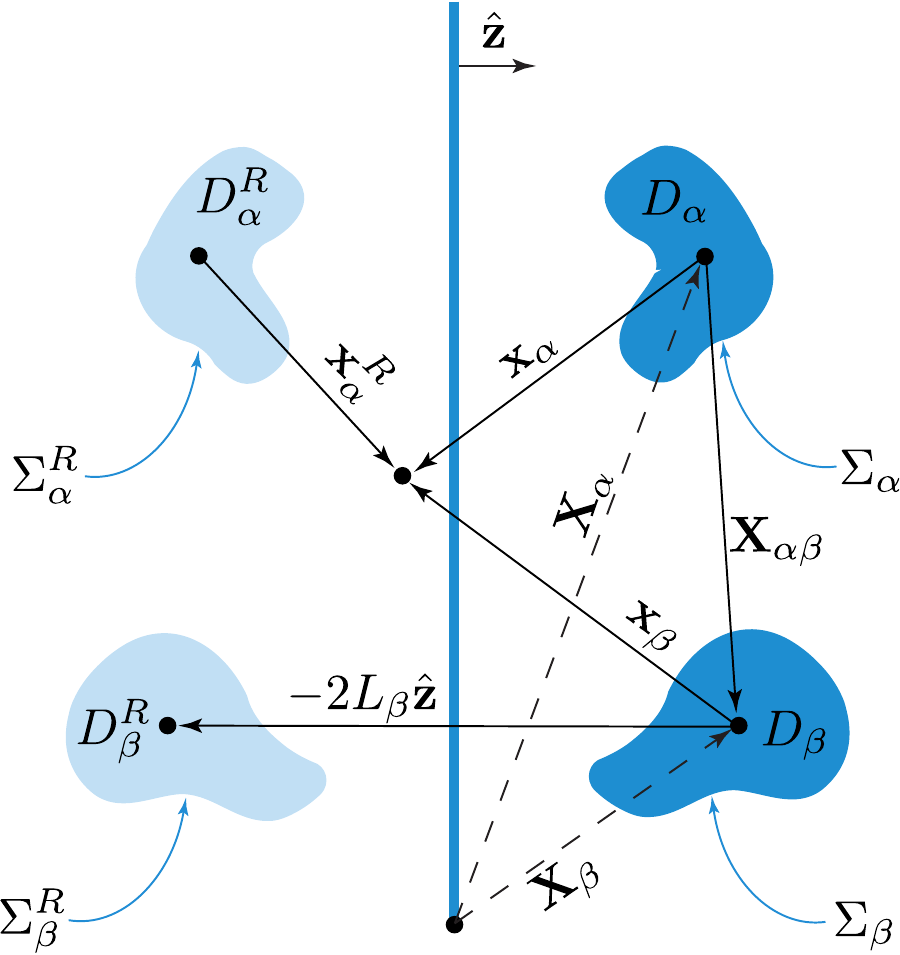}
\caption{Geometry with original surfaces $\Sigma_\alpha$ (objects
  $D_\alpha$) in the right half space ($z>0$) and mirror surfaces
  $\Sigma_\alpha^R$ (objects $D_\alpha^R$).  $\bfx_\alpha$,
  $\bfx_\alpha^R$ are local coordinate vectors measured from the
  object's origins, $L_\beta$ is the center-to-mirror distance
  of surface $\Sigma_\beta$ (object $D_\beta$).}
\label{fig:objects}
\end{figure}

The surfaces are fixed and the boundary conditions are time
independent.  Hence each Fourier component of the field with respect
to $x^0$ obeys the constraints at the surfaces separately. To make use
of this property, we expand $\phi(\bfx,x^0)$ as
\begin{equation}
  \label{eq:Fourier-decomp}
  \phi(\bfx,x^0) = \sum_{n=-\infty}^\infty \phi_n(\bfx) e^{i \kappa_n x^0}
\end{equation}
with Matsubara frequencies $\kappa_n=2\pi n/\Lambda$ and
$\phi_{-n}(\bfx)=\phi^*_n(\bfx)$. The functional integral splits
into independent functional integrals over the $\phi_n(\bfx)$ and
the logarithm of $Z$ can be written as the sum
\begin{eqnarray}
  \label{eq:log-Z-as-sum}
 && \ln Z = \\
&& \!\!\!\sum_{n=-\infty}^\infty \!\ln \int [\cD \phi_n]_\cC
\exp \left[-\frac{\Lambda}{2\hbar c} \int_> \!\!d\bfx \left\{
\kappa_n^2 |\phi_n|^2 + |\nabla \phi_n|^2
\right\}\right] \nonumber\, .
\end{eqnarray}
In the following we are interested in the limit of zero temperature.
Then $\Lambda\to\infty$, and the sum over $n$ can be replaced by the
integral $\frac{\Lambda}{2\pi} \int_{-\infty}^\infty d\kappa$ and $\phi_n(\bfx)$ is replaced
by $\phi(\bfx,\kappa)$. Combining positive and negative $\kappa$, we get 
\begin{equation}
  \label{eq:Z-of-k}
  \ln Z = \frac{\Lambda}{2\pi} \int_0^\infty d\kappa \ln \cZ(\kappa)
\end{equation}
with
\begin{widetext}
\begin{equation}
  \label{eq:log-Z-as-integral}
 \cZ(\kappa) \!=\! \int [\cD \phi(\bfx,\kappa)]_\cC[\cD \phi^*(\bfx,\kappa)]_\cC
\exp \left[-\frac{\Lambda}{\hbar c} \int_> \!\!d\bfx \left\{
\kappa^2 |\phi|^2 + |\nabla \phi|^2
\right\}\right] \, .
\end{equation}
\end{widetext}
The ground state energy is obtained from $Z$ as
\begin{eqnarray}
  \label{eq:gs-energy}
  \cE_0 &=& - \lim_{\beta\to\infty} \frac{1}{\beta} \ln Z\nonumber\\
&=& -\frac{\hbar c}{2\pi} \int_0^\infty d\kappa \ln \cZ(\kappa) \, .
\end{eqnarray}
The Casimir energy is the difference between the ground state energy
of the surfaces at their actual distance and the sum of the ground
state energies of the separate surfaces which is obtained by removing
the surfaces to infinite separation,
\begin{equation}
  \label{eq:casimir-energy}
  \cE = -\frac{\hbar c}{2\pi} \int_0^\infty d\kappa \ln 
\frac{\cZ(\kappa)}{\cZ_\infty(\kappa)} \, ,
\end{equation}
where $\cZ_\infty(\kappa)$ is the partition function for the surfaces
with infinite separation.  In the following we suppress the label
$\kappa$.  Next, the constraints at the objects are implemented by
functional $\delta$-functions \cite{Bordag+85,LK91}. For Dirichlet
boundary conditions, $\phi=0$, on the surfaces $\Sigma_\alpha$, the
constraint functional integral can be expressed in terms of an
unconstrained integral by using
\begin{eqnarray}
  \label{eq:func-delta-fcts}
&&  \int [\cD \phi]_\cC [\cD \phi^*]_\cC = \\
&&\!\!\!\!\!\int \!\!\cD \phi \cD \phi^*
\prod_{\alpha=1}^N \int \!\!\cD \varrho_\alpha \cD \varrho^*_\alpha
\exp\left[ i\!\!\int_{\Sigma_\alpha}\!\!\!\!\! d\bfx \,\left\{\varrho_\alpha^*(\bfx) \phi(\bfx) +\cc
\right\} \right]  ,\nonumber
\end{eqnarray}
where the $\delta$-functions at each position of the surfaces have been
written as an integral over a source field $\varrho_\alpha(\bfx)$ that
is non-zero on the surfaces $\Sigma_\alpha$ only. When we use this
representation of the constraints in Eq.~\eqref{eq:log-Z-as-integral}, the
now unconstrained functional integral over $\phi(\bfx)$ is Gaussian and
yields
\begin{widetext}
\begin{eqnarray}
  \label{eq:Z-of-kappa-sources}
\cZ(\kappa)&=&\cZ_0 \prod_{\alpha=1}^N\int \cD\varrho_\alpha \cD\varrho^*_\alpha\exp\left[
-\frac{\hbar c}{2\Lambda} \sum_{\alpha\beta}\int_{\Sigma_\alpha} \!\! 
d\bfx \int_{\Sigma_\beta} \!\!d\bfx' 
\left\{\varrho^*_\alpha(\bfx) G_>(\bfx,\bfx',\kappa) \varrho_\beta(\bfx') +\cc\right\}
\right] \nonumber\\
&\equiv& \cZ_0 \prod_{\alpha=1}^N\int \cD\varrho_\alpha\cD\varrho^*_\alpha
e^{-(\hbar c/\Lambda) \tilde S[\varrho]}\, ,
\end{eqnarray}
\end{widetext}
where $\cZ_0$ is the partition function of the field in the right half
space without the surfaces $\Sigma_\alpha$. For Neumann boundary
conditions at the surfaces $\Sigma_\alpha$ the field $\phi(\bfx)$ in
the exponential of Eq.~\eqref{eq:func-delta-fcts} is replaced by
$\partial_n \phi(\bfx)$ and $G_>(\bfx,\bfx',\kappa)$ in
Eq.~\eqref{eq:Z-of-kappa-sources} is replaced by
$\partial_n\partial_{n'}G_>(\bfx,\bfx',\kappa)$, where $\partial_n$ is
the normal derivative pointing out of the objects.  Here
$G_>(\bfx,\bfx',\kappa)$ is the free Green's function in the right half
space which is given by
\begin{equation}
  \label{eq:free_G_D}
  G_>(\bfx,\bfx',\kappa) = G_0(\bfx,\bfx',\kappa) \mp G_0(\bfx,\bfx_R',\kappa) \, ,
\end{equation}
where $\bfx_R=(\bfx_\|,-z)$ is the mirror image of $\bfx=(\bfx_\|,z)$
and the $-$($+$) sign applies to Dirichlet (Neumann) boundary conditions at
the mirror plane at $z=0$. The Green's function of free unbounded space
is given by
\begin{equation}
  \label{eq:G_free}
  G_0(\bfx,\bfx',\kappa) = \frac{e^{ik|\bfx-\bfx'|}}{4\pi|\bfx-\bfx'|}_{|k\to i\kappa} \, .
\end{equation}
Using Eq.~\eqref{eq:free_G_D} the action $\tilde S[\varrho]$ defined by
Eq.~\eqref{eq:Z-of-kappa-sources} can be expressed in terms of the original sources
$\varrho_\alpha(\bfx)$ and their mirror sources
$\varrho^R_\alpha(\bfx)$,
\begin{eqnarray}
  \label{eq:Z-of-mirror-sources}
 && \tilde S[\varrho] = \frac{1}{2} \sum_{\alpha\beta} \left\{ 
\int_{\Sigma_\alpha} d\bfx\int_{\Sigma_\beta} d\bfx' \varrho^*_\alpha(\bfx) G_0(\bfx,\bfx',\kappa)
\varrho_\beta(\bfx') \right. \nonumber\\
&+&\!\!\! \left.  
\int_{\Sigma_\alpha} d\bfx\int_{\Sigma^R_\beta} d\bfx' \varrho^*_\alpha(\bfx) G_0(\bfx,\bfx',\kappa)
\varrho^R_\beta(\bfx') +\cc \right\} 
\end{eqnarray}
for Dirichlet boundary conditions at the surfaces $\Sigma_\alpha$
and with $G_0(\bfx,\bfx',\kappa)$ replaced by
$\partial_n \partial_{n'} G_0(\bfx,\bfx',\kappa)$ for Neumann
boundary conditions at the surfaces $\Sigma_\alpha$. Here we have
introduced the mirror sources 
\begin{eqnarray}
  \label{eq:mirror-sources}
  \varrho_\alpha^R(\bfx) = \mp \varrho_\alpha(\bfx_R) \, ,
\end{eqnarray}
where the $-$ ($+$) sign applies to a Dirichlet (Neumann) mirror.  The
mirror sources are located on the mirror surfaces $\Sigma_\alpha^R$
that are obtained from the $\Sigma_\alpha$ by $z\to -z$ for all
surface positions, see Fig.~\ref{fig:objects}. The first term of the
action of Eq.~\eqref{eq:Z-of-mirror-sources} describes the interaction
of the surface sources in the absence of the mirror plane. The second
term couples each surface source to all mirror sources. Since the
mirror problem is now described by an action in {\it free} space with
sources and mirror sources, we can apply the concepts of the previously
developed approach for Casimir interactions between compact objects in
unbounded space \cite{Emig+07}. Below, we provide an explicit
derivation for the case of Dirichlet boundary conditions at the
surfaces $\Sigma_\alpha$ but we shall also indicate how the derivation
has to be modified for Neumann boundary conditions.

The action of Eq.~\eqref{eq:Z-of-mirror-sources} is composed of two
qualitatively different terms that we will now consider
separately. Firstly, there are terms that couple sources on different
surfaces ($\bfx \neq \bfx'$) where we use the term ``surface'' in the
following for the original and the mirror surfaces. We shall call
these terms {\it off-diagonal}. As {\it diagonal} terms we shall
denote those which couple sources on the same surface ($\bfx = \bfx'$
possible). Both type of terms can be expressed in terms of 
the multipole moments of the sources. 

{\it Off-diagonal terms} --- These terms couple sources on
different objects,
\begin{equation}
  \label{eq:S-diag-orig}
  \tilde S_{\alpha\beta} = \frac{1}{2} \int_{\Sigma_\alpha} d\bfx_\alpha \left\{
\varrho^*_\alpha(\bfx_\alpha)
\phi_\beta(\bfx_\alpha) +\cc\right\}
\end{equation}
for $\alpha\neq\beta$ and the original sources to the mirror sources,
\begin{equation}
  \label{eq:S-diag-mirror}
   \tilde S^R_{\alpha\beta} = \frac{1}{2} \int_{\Sigma_\alpha} d\bfx_\alpha 
\left\{ \varrho^*_\alpha(\bfx_\alpha)
\phi^R_\beta(\bfx_\alpha)+\cc\right\} 
\end{equation}
for {\it all} $\alpha$, $\beta$. Here we have introduced local coordinates
$\bfx_\alpha$ that are measured relative to an arbitrarily chosen
origin that is located inside the surface $\alpha$, see Fig.~\ref{fig:objects}.
We have also defined the fields
\begin{eqnarray}
  \label{eq:phi_class}
  \phi_\beta(\bfx) &=& \int_{\Sigma_\beta} d\bfx' G_0(\bfx,\bfx',\kappa) \varrho_\beta(\bfx') 
\nonumber\\
  \phi^R_\beta(\bfx) &=& \int_{\Sigma^R_\beta} d\bfx' G_0(\bfx,\bfx',\kappa) 
\varrho^R_\beta(\bfx') \, , 
\end{eqnarray}
which are the classical fields generated by the sources.  For Neumann
boundary conditions on the surfaces $\Sigma_\alpha$,
$\phi_\beta(\bfx_\alpha)$ and $\phi^R_\beta(\bfx_\alpha)$ in
Eqs.~\eqref{eq:S-diag-orig} and \eqref{eq:S-diag-mirror} have to be
replaced by $\partial_n \phi_\beta(\bfx_\alpha)$ and $\partial_n
\phi^R_\beta(\bfx_\alpha)$, respectively. Also,
$G_0(\bfx,\bfx',\kappa)$ in Eq.~\eqref{eq:phi_class} has to be
replaced by $\partial_{n'}G_0(\bfx,\bfx',\kappa)$.

Since we can assume that every position on $\Sigma_\alpha$ is outside
a sphere enclosing $\Sigma_\beta$ or $\Sigma^R_\beta$, we can use the
partial wave expansion of $G_0(\bfx,\bfx',\kappa)$ for $|\bfx'| <
|\bfx|$,
\begin{equation}
  \label{eq:G-partial-waves}
  G_0(\bfx,\bfx',\kappa) = -\kappa \sum_{lm} j_l(i\kappa r') h_l^{(1)}(i\kappa r)
Y_{lm}(\hat\bfx) Y_{lm}^*(\hat\bfx') \, .
\end{equation}
When we consider Eq.~\eqref{eq:phi_class} with coordinates
$\bfx_\beta$ relative to the origin inside surface $\Sigma_\beta$, the field
that is generated by the source $\varrho_\beta$ can be written as
\begin{equation}
  \label{eq:phi-of-mps}
  \phi_\beta(\bfx_\beta) = -\kappa \sum_{lm} Q_{\beta,lm} h_l^{(1)} (i\kappa r_\beta)
Y_{lm}(\hat\bfx_\beta) \, ,
\end{equation}
where we have defined the multipole moments of the source $\varrho_\beta$ as
\begin{equation}
  \label{eq:def-multipoles}
  Q_{\beta,lm}= \int_{\Sigma_\beta} d\bfx_\beta j_l(i\kappa r_\beta) Y_{lm}^*(\hat\bfx_\beta)
\varrho_\beta(\bfx_\beta) \, . 
\end{equation}
The field of the mirror sources can be expressed in the same form,
\begin{equation}
  \label{eq:phi-of-mps-mirror}
  \phi^R_\beta(\bfx^R_\beta) = -\kappa \sum_{lm} Q^R_{\beta,lm} h_l^{(1)} (i\kappa r^R_\beta)
Y_{lm}(\hat\bfx^R_\beta) \, ,
\end{equation}
where $\bfx^R_\beta$ denotes the local coordinates that are measured
relative to the origin inside surface $\Sigma_\beta^R$. The multipole
moments of the mirror source are given by
\begin{equation}
  \label{eq:mirror-multipoles}
  Q_{\beta,lm}^R = \mp (-1)^{l+m} Q_{\beta,lm}
\end{equation}
due to Eq.~\eqref{eq:mirror-sources} and
$Y_{lm}(\hat\bfx_R)=(-1)^{l+m}Y_{lm}(\hat\bfx)$. For Neumann boundary
conditions at the surfaces $\Sigma_\alpha$ the above expressions
remain valid if the multipole moments $Q_{\beta,lm}$ are replaced by
Neumann multipoles which have the form of
Eq.~\eqref{eq:def-multipoles} but with $j_l(i\kappa r_\beta) Y_{lm}^*(\hat\bfx_\beta)$
replaced by $\partial_n [ j_l(i\kappa r_\beta) Y_{lm}^*(\hat\bfx_\beta)]$.

In order to express the action of Eq.~\eqref{eq:S-diag-orig} in terms
of multipole moments, we have to write the field generated by the
surface source $\varrho_\beta$ as a function of the local coordinate
$\bfx_\alpha$ that is regular at the origin inside $\Sigma_\alpha$.
This can be done using translation matrices ${\mathbb U}$ which relate
outgoing ($h_l^{(1)}$) and regular ($j_l$) spherical Bessel functions,
\begin{equation}
  \label{eq:def-transl-matrix}
  h^{(1)}_{l}(i\kappa r_{\b})Y_{lm}(\hat\bfx_{\b}) =
\sum_{l'm'}\cU_{l'm'lm}(\bfX_{\a\b})
j_{l'}(i\kappa r_{\a})Y_{l'm'}(\hat\bfx_{\a}) \, .
\end{equation}
The matrix elements $\cU_{l'm'lm}$ of ${\mathbb U}$ depend on the
vector
$\bfX_{\alpha\beta}=\bfX_\beta-\bfX_\alpha=\bfx_\alpha-\bfx_\beta$
from the origin inside $\Sigma_\alpha$ to the origin inside
$\Sigma_\beta$, see Fig.~\ref{fig:objects}, and are given by \cite{Emig+07}
\begin{widetext}
\begin{eqnarray}
  \label{eq:U-matrix-elements}
\cU_{l'm'lm}(\bfX_{\a\b}) &=&
\sqrt{4\pi} (-1)^{m} i^{l-l'}\sqrt{(2l+1)(2l'+1)} \nonumber\\
&\times& \sum_{l''}  i^{l''} \sqrt{2l''+1}
\begin{pmatrix}l&l'&l''\\0&0&0\end{pmatrix}
\begin{pmatrix}l&l'&l''\\m&-m'&m'-m\end{pmatrix}
h_{l''}^{(1)}(i\kappa |\bfX_{\a\b}|)Y_{l''m-m'}(\hat\bfX_{\a\b}) \, ,
\end{eqnarray}
\end{widetext}
where we have assumed that the Cartesian coordinate frames associated
with the two origins have identical orientation, i.e., they are
related by a translation. The summation over $l''$ involves only a
finite number of terms since the 3-$j$ symbols vanish for $l''>l+l'$
and $l''<|l-l'|$. 

Using the translation formula of Eq.~\eqref{eq:def-transl-matrix}, the
field generated by the source on surface $\Sigma_\beta$, given by
Eq.~\eqref{eq:phi-of-mps}, can be written as function of the coordinate
$\bfx_\alpha$ as
\begin{equation}
  \label{eq:phi-translated}
  \phi_\beta(\bfx_\alpha) = -\kappa \sum_{lm} Q_{\beta,lm} \sum_{l'm'} 
\cU_{l'm'lm}^{\alpha\beta} j_{l'}(i \kappa r_\alpha) Y_{l'm'}(\hat\bfx_\alpha) \, ,
\end{equation}
where we have introduced $\cU_{l'm'lm}^{\alpha\beta} \equiv
\cU_{l'm'lm}(\bfX_{\alpha\beta})$. Similarly we obtain for the field
generated by the mirror sources of Eq.~\eqref{eq:phi-of-mps-mirror},
now expressed as function of $\bfx_\alpha$, 
\begin{equation}
  \label{eq:phi-mirror-translated}
    \phi^R_\beta(\bfx_\alpha) = \pm\kappa \sum_{lm} Q_{\beta,lm}\sum_{l'm'} 
\cU_{l'm'lm}^{R,\alpha\beta} j_{l'}(i \kappa r_\alpha) Y_{l'm'}(\hat\bfx_\alpha) \, ,
\end{equation}
where we defined
\begin{equation}
\label{eq:U-matrix-elements-R}
\cU_{l'm'lm}^{R,\alpha\beta} \equiv
(-1)^{l+m} \cU_{l'm'lm}(\bfx_\alpha-\bfx^R_\beta) \, ,
\end{equation}
and used Eq.~\eqref{eq:mirror-multipoles}.  Notice that the latter
formula applies also to the case $\alpha=\beta$ which describes the
translation between the surface $\Sigma_\alpha$ and its mirror image
so that the argument of the translation matrix becomes
$\bfx_\alpha-\bfx_\alpha^R= -2L_\alpha\hat\bfz$ where $L_\alpha$ is
the normal distance between the origin of surface $\Sigma_\alpha$ and
the mirror plane. For this case of translations along the
$\hat\bfz$-direction the translation matrix simplifies to
\cite{Emig+07}
\begin{widetext}
\begin{equation}
  \label{eq:U-matrix-elements-z-direction}
\cU_{l'm'lm}(-2L_\alpha\hat\bfz) =\delta_{m'm}
 (-1)^{m} i^{l-l'}\sqrt{(2l+1)(2l'+1)} 
\sum_{l''}  i^{-l''} (2l''+1)
\begin{pmatrix}l&l'&l''\\0&0&0\end{pmatrix}
\begin{pmatrix}l&l'&l''\\m&-m&0\end{pmatrix}
h_{l''}^{(1)}(i\kappa 2L_\alpha) \, .
\end{equation}
\end{widetext}
When we substitute the result for the fields of
Eqs.~\eqref{eq:phi-translated}, \eqref{eq:phi-mirror-translated} into
Eqs.~\eqref{eq:S-diag-orig}, \eqref{eq:S-diag-mirror}, we 
obtain the action in terms of the original multipole moments,
\begin{eqnarray}
  \sum_{\alpha\neq\beta}\tilde S_{\alpha\beta} &=& -\kappa \sum_{\alpha\neq\beta}
\sum_{lml'm'} Q^*_{\alpha,l'm'} 
\,\,\tilde \cU^{\alpha\beta}_{l'm'lm}\, Q_{\beta,lm}\nonumber \\
  \label{eq:S-off-diag-of-Qs}\\
 \sum_{\alpha\beta} \tilde S^R_{\alpha\beta} &=&\pm \kappa \sum_{\alpha\beta}
\sum_{lml'm'} Q^*_{\alpha,l'm'} 
\,\,\tilde \cU^{R,\alpha\beta}_{l'm'lm}\, Q_{\beta,lm} \, ,\nonumber\\
  \label{eq:S-off-diag-of-Qs-R}
\end{eqnarray}
which applies to Dirichlet as well as Neumann boundary conditions at
the surfaces $\Sigma_\alpha$.  To simplify notation, we have defined
the modified translation matrix elements
\begin{eqnarray}
  \label{eq:transl-mat-mod}
  \tilde\cU^{\alpha\beta}_{l'm'lm} &\equiv& (-1)^{l'}\cU^{\alpha\beta}_{l'm'lm} \\
  \label{eq:transl-mat-mod-R-diag}
  \tilde\cU^{R,\alpha\alpha}_{l'mlm} &\equiv& (-1)^{l'}\cU^{R,\alpha\alpha}_{l'mlm} \\
  \label{eq:transl-mat-mod-R}
  \tilde\cU^{R,\alpha\beta}_{l'm'lm} &\equiv& \frac{1}{2}\left[
(-1)^{l'}\cU^{R,\alpha\beta}_{l'm'lm} + (-1)^l\cU^{{R,\beta\alpha}^*}_{lml'm'} \right],\,
(\alpha\neq\beta) \nonumber\\
\end{eqnarray}
Here we have used the definition of Eq.~\eqref{eq:def-multipoles} and
$[j_l(i\kappa r)]^*=(-1)^lj_l(i\kappa r)$. In addition, we applied the
symmetry relations $\cU^{{\alpha\beta}^*}_{lml'm'} = (-1)^{l+l'}
\cU^{\beta\alpha}_{l'm'lm}$ and $\cU^{{R,\alpha\alpha}^*}_{lml'm} =
(-1)^{l+l'} \cU^{R,\alpha\alpha}_{l'mlm}$ which follow from
Eqs.~\eqref{eq:U-matrix-elements},
\eqref{eq:U-matrix-elements-z-direction}, symmetry properties of 3-j
symbols and spherical harmonics, and $h_l^{(1)}(iz)=-i^{-l}
\sqrt{2/(\pi z)} K_{l+1/2}(z)$ where $K_{l+1/2}(z)$ is the (real
valued) modified Bessel function of second kind.  Notice that the
actions of Eqs.~\eqref{eq:S-off-diag-of-Qs},
\eqref{eq:S-off-diag-of-Qs-R} couple sources on different surfaces and
hide the dependence on the particular boundary conditions and shape of
the surfaces.

{\it Diagonal terms} --- These are the self-action terms 
\begin{equation}
  \label{eq:S-off-diag}
  \tilde S_\alpha  = \frac{1}{2} \int_{\Sigma_\alpha} d\bfx
\left\{ \varrho^*_\alpha(\bfx) \phi_\alpha(\bfx) + \cc \right\}
\end{equation}
in Eq.~\eqref{eq:Z-of-mirror-sources}, which we have expressed here in
terms of the classical field defined in Eq.~\eqref{eq:phi_class}.  For
Neumann boundary conditions on the surface $\Sigma_\alpha$ we have to
replace again $\phi_\alpha(\bfx)$ by $\partial_n \phi_\alpha(\bfx)$ in
Eq.~\eqref{eq:S-off-diag}.  Here and in the following we only use the
coordinate system associated with the origin inside $\Sigma_\alpha$,
and hence drop the label $\alpha$ on the coordinates. The classical
field generated by the source $\varrho_\alpha(\bfx)$ on surface
$\Sigma_\alpha$ obeys the Helmholtz equation
\begin{equation}
  \label{eq:Helmholtz-class-field}
  (\nabla^2-\kappa^2) \phi_\alpha(\bfx)= -\varrho_\alpha(\bfx) \, .
\end{equation}
For positions $\bfx$ that are located on the surface $\Sigma_\alpha$
the {\it total} field $\phi(\bfx)$ generated by all sources must obey
the same Helmholtz equation. The part of the total field that is
generated by sources other than $\varrho_\alpha(\bfx)$ can be regarded
as incident field $\phi_{0,\alpha}(\bfx)$ at the surface
$\Sigma_\alpha$, which obeys in region around $\Sigma_\alpha$, that is
free of sources other than $\varrho_\alpha(\bfx)$, the homogeneous
Helmholtz equation.  Hence the total field can be written as
\begin{eqnarray}
  \label{eq:total-class-field}
  \phi(\bfx) &=&\phi_{0,\alpha}(\bfx) +\phi_\alpha(\bfx)\\
&=& \phi_{0,\alpha}(\bfx) + \int_{\Sigma_\alpha} d\bfx' \,G_0(\bfx,\bfx',\kappa)
\varrho_\alpha(\bfx') \nonumber
\end{eqnarray}
for all $\bfx$ located on $\Sigma_\alpha$. For Neumann boundary
conditions at $\Sigma_\alpha$ the Green's function in
Eq.~\eqref{eq:total-class-field} is replaced again by $\partial_{n'}
G_0(\bfx,\bfx',\kappa)$.  We would like to evaluate the action of
Eq.~\eqref{eq:S-off-diag} in terms of multipole moments. Hence, we
must consider field configurations with a fixed source on surface
$\Sigma_\alpha$ that is characterized by its multipole moments. This
implies that we have to find the incident field
$\phi_{0,\alpha}(\bfx)$ that induces a prescribed set of multipoles
$Q_{\alpha,lm}$ on $\Sigma_\alpha$. The multipole moments can be
identified as the amplitudes of the scattered field which is given by
$\phi_\alpha(\bfx)$ with $\bfx$ located outside of the surface
$\Sigma_\alpha$. Using the partial wave expansion of
Eq.~\eqref{eq:G-partial-waves}, we get
\begin{equation}
  \label{eq:phi-scattered}
  \phi_\alpha(\bfx)=-\kappa \sum_{l'm'} Q_{\alpha,l'm'} h_{l'}^{(1)}(i\kappa r)
Y_{l'm'}(\hat\bfx) \, . 
\end{equation}
From scattering theory we know that the amplitudes of the scattered
field are related to the amplitudes of the regular incident field by the
transition matrix ${\mathbb T}$ which is related to the
scattering matrix ${\mathbb S}$ by ${\mathbb T}=({\mathbb S}-1)/2$.
If we expand the incident field as
\begin{equation}
  \label{eq:exp-incident-field}
  \phi_{0,\alpha}(\bfx)= \sum_{lm} \phi_{0,\alpha,lm} \, j_l(i\kappa r)Y_{lm}(\hat\bfx) \, , 
\end{equation}
the amplitudes of the scattered field are given by
\begin{equation}
  \label{eq:scatt-amp}
  -\kappa Q_{\alpha,l'm'} = \sum_{lm} \cT^\alpha_{l'm'lm}\, \phi_{0,\alpha,lm} \, , 
\end{equation}
where the $\cT^\alpha_{l'm'lm}$ denote the matrix elements of the ${\mathbb T}$-matrix
of the surface $\Sigma_\alpha$. Hence, the amplitudes of the incident field
have to be given by
\begin{equation}
  \label{eq:inc_amp}
  \phi_{0,\alpha,lm} = -\kappa \sum_{l'm'} [\cT^\alpha]^{-1}_{lml'm'} Q_{\alpha,l'm'} \, .
\end{equation}
For Dirichlet boundary conditions, the total field $\phi(\bfx)$ of
Eq.~\eqref{eq:total-class-field} has to vanish on $\Sigma_\alpha$ so
that $\phi_\alpha(\bfx)=-\phi_{0,\alpha}(\bfx)$ on the surface. Hence,
using Eqs.~\eqref{eq:exp-incident-field} and \eqref{eq:inc_amp}, the action
of Eq.~\eqref{eq:S-off-diag} can be expressed in terms of the multipole
moments of $\varrho_\alpha(\bfx)$,
\begin{equation}
  \label{eq:S-diag-of-Qs}
  \tilde S_\alpha = \kappa \sum_{lml'm'} Q^*_{\alpha,lm}\, [\tilde \cT^\alpha]^{-1}_{lml'm'}
\,Q_{\alpha,l'm'}
\end{equation}
with the modified $\tilde{\mathbb T}$-matrix defined by
\begin{equation}
  \label{eq:T-mod}
\tilde  \cT^\alpha_{l'm'lm} \equiv (-1)^l \cT^\alpha_{l'm'lm} \, .
\end{equation}
Here we have used the definition of the multipoles in
Eq.~\eqref{eq:def-multipoles} to integrate over the surface and 
applied the relation $\cT^{\alpha^*}_{l'm'lm}=(-1)^{l+l'}\cT^\alpha_{lml'm'}$.

For Neumann boundary conditions on $\Sigma_\alpha$, the normal
derivative of the total field of Eq.~\eqref{eq:total-class-field} has
to vanish on $\Sigma_\alpha$ so that $\partial_n
\phi_\alpha(\bfx)=-\partial_n \phi_{0,\alpha}(\bfx)$ on the surface. 
When we use the definition of the multipole moments for Neumann
boundary conditions, we obtain again Eq.~\eqref{eq:S-diag-of-Qs} but
with the matrix ${\mathbb T}^\alpha$ for Neumann boundary conditions.

Now we can combine all results for off-diagonal and diagonal terms and
express the total action of Eq.~\eqref{eq:Z-of-mirror-sources} in
terms of the original multipole moments. Since we have
\begin{equation}
  \label{eq:action-as-sum-of-3}
  \tilde S = \sum_{\alpha\neq\beta} \tilde S_{\alpha\beta} + \sum_\alpha \tilde S_\alpha
+ \sum_{\alpha,\beta} \tilde S^R_{\alpha\beta}
\end{equation}
we obtain from Eqs.~\eqref{eq:S-off-diag-of-Qs}, \eqref{eq:S-off-diag-of-Qs-R} and
\eqref{eq:S-diag-of-Qs} the total action of the multipole
moments
\begin{equation}
  \label{eq:total_action-final}
  \tilde S[Q] = \sum_{\alpha\beta} Q^*_\alpha {\mathbb M}^{\alpha\beta} Q_\beta \, ,
\end{equation}
where we have suppressed the sum over the indices $l$,
$m$, $l'$, $m'$ and defined the matrix
\begin{equation}
  \label{eq:def-matrix-M}
  {\mathbb M}^{\alpha\beta}= \kappa \left\{ [\tilde {\mathbb T}^\alpha ]^{-1} 
\delta_{\alpha\beta} - \tilde {\mathbb U}^{\alpha\beta} (1-\delta_{\alpha\beta})
\pm \tilde {\mathbb U}^{R,\alpha\beta} \right\} 
\end{equation}
for Dirichlet ($+$) or Neumann ($-$) boundary conditions at the mirror
plane.  The partition function of Eq.~\eqref{eq:Z-of-kappa-sources} is
then obtained by integrating over all multipole moments,
\begin{equation}
  \label{eq:Z-og-kappa-final}
  \cZ(\kappa)=\cZ_0 \prod_{\alpha=1}^N \int \! d Q_\alpha d Q^*_\alpha \exp \left[ 
-\frac{\hbar c}{\Lambda}
\sum_{\alpha\beta} Q^*_\alpha {\mathbb M}^{\alpha\beta} Q_\beta 
\right] \, .
\end{equation}
The Gaussian integral over the multipoles is proportional to 
the inverse determinant of ${\mathbb M}^{\alpha\beta}$.
Finally, we substitute into Eq.~\eqref{eq:casimir-energy} to obtain
the Casimir energy,
\begin{equation}
  \label{eq:Casimir-energy-gen-result}
  \cE = \frac{\hbar c}{2\pi} \int_0^\infty d\kappa \ln\frac{\det {\mathbb M}}
{\det {\mathbb M}_\infty} \, ,
\end{equation}
where the determinant is taken with respect to the partial wave
indices $l$, $m$ and the surface indices $\alpha$, $\beta$.  The
matrix ${\mathbb M}_\infty$ is the result of moving the surfaces to
infinite separation, where the translation matrices $\tilde{\mathbb
  U}$, $\tilde{\mathbb U}^R$ vanish so that 
\begin{equation}
  \label{eq:M-infty}
  {\mathbb M}^{\alpha\beta}_\infty =  \kappa \, [\tilde {\mathbb T}^\alpha ]^{-1} 
\delta_{\alpha\beta} \, .
\end{equation}
In the special case of one compact surface in front of the mirror plane
Eq.~\eqref{eq:Casimir-energy-gen-result} simplifies to
\begin{equation}
  \label{eq:casimir-energy-1-surface}
  \cE_1 = \frac{\hbar c}{2\pi} \int_0^\infty d\kappa \ln \det \left(
1 \pm {\mathbb T}^1 {\mathbb U}^{R,11} \right)
\end{equation}
for Dirichlet (+) or Neumann (-) boundary conditions at the mirror
plane. This expression applies to Dirichlet, Neumann and even more
general boundary conditions at the compact surface which enter only
via the matrix ${\mathbb T}^1$.  Notice that this result depends on
the original matrix elements (without tilde) since the phase factors of
Eqs.~\eqref{eq:transl-mat-mod-R-diag} and \eqref{eq:T-mod} drop out when
taking the matrix product of $\tilde {\mathbb T}^1$ and $\tilde
{\mathbb U}^{R,11}$.  This general result shows that the Casimir
interaction between a mirror and an object with arbitrary shape and
boundary condition can be obtained from the transition matrix
${\mathbb T}^1$ of the object and the translation matrix ${\mathbb
  U}^{R,11}$ that describes the (classical) interaction
between the induced source and its mirror image.

\subsection{Electromagnetic field}

The derivation of the Casimir energy for a scalar field can be extended
to electromagnetic field fluctuations in the presence of dielectric
objects \cite{Emig+07b}. The result will have the form of an effective
action for electric and magnetic multipoles of the current densities
$\bJ_\alpha$ inside the objects. We consider again $N$ objects
that are located in the right half space that is bounded by a
perfectly conducting plane at $z=0$. At his plane the tangential
electric field and the normal magnetic field vanish, $\bE_\|=0$,
$\bB_\perp=0$. The material objects are assumed to be dielectrics that
are characterized by a frequency dependent dielectric function
$\epsilon_\alpha(\omega)$ and permeability function
$\mu_\alpha(\omega)$. 

The partition function can be factorized again into a product of factors
$\cZ(\kappa)$ at a fixed Wick rotated frequency $\kappa$. Hence, in the
following we consider all expressions at fixed $\kappa$ and suppress
the label $\kappa$.  The Euclidean action for the electromagnetic field in
the presence of macroscopic media without external sources can be
expressed as
\begin{equation}
  \label{eq:EM-S-macro}
  S_0 = \frac{1}{2} \int_> d\bfx  \left[ \bE \bD^* + \bB \bH^* +\cc \right] 
\end{equation}
in terms of the macroscopic fields $\bD=\epsilon(i\kappa)\bE$,
$\bH=\mu^{-1}(i\kappa)\bB$. Here the {\it energy} density of the field
is integrated since under a Wick rotation to imaginary time the
Lagrangian in real time is generally transformed to the Hamiltonian in
imaginary time.  In this description the induced (bound) currents
inside the material objects have been absorbed into the definition of
the macroscopic fields. The partition function for this action is
given by a functional integral over the vector potential $\bA$ and the
scalar potential $\Phi$ (after introducing a Faddeev-Popov gauge
fixing term) where the fields are expressed in terms of the potentials
as
\begin{equation}
  \label{eq:gauge-potentials}
  \bE = i\kappa \bA -\nabla \Phi, \quad \bB= \nabla \times \bA \, .
\end{equation}

An alternative description in terms of the fields $\bE$, $\bB$ only is
obtained if the bound charges ($\rho$) and currents ($\bJ$) density
inside the objects are not substituted by the macroscopic fields but
considered explicitly. Then the Wick rotated action can be written in
terms of the potentials as
\begin{eqnarray}
  \label{eq:EM-S-w-sources}
  S[\bJ,\rho] &=&  \int_> d\bfx \bigg[ \kappa^2 (|\bA|^2+|\Phi|^2 ) 
+\sum_{j=1}^3 |\nabla A_j|^2 +|\nabla\Phi|^2 \nonumber\\
&+& \left(\bA \bJ^* + \Phi \rho^* +\cc \right)\bigg] \, ,
\end{eqnarray}
where we have chosen the Feynman gauge.  The partition function is
obtained by integrating over both potentials $\bA$, $\Phi$ and sources
$\bJ$, $\rho$.  However, in the latter integration the currents and charges
must be weighted according to the energy cost for inducing them on the objects.
This energy cost must depend on shape and material of the
objects.  We will see below that this can be achieved by rewriting the
self-energies of the separate objects in terms of the incident field
that generates the polarizations and magnetizations which give rise
to the induced current.  Hence the action of
Eq.~\eqref{eq:EM-S-w-sources} is independent of material and shape of
the objects and these properties enter the partition function
through proper weights on the currents that measure the susceptibility
of the objects to current fluctuations.

We proceed by integrating out the unconstrained fluctuations of the
potentials $\bA$, $\Phi$ in the action of
Eq.~\eqref{eq:EM-S-w-sources}.  This integration yields the partition
function as a {\it weighted} functional integral over sources which
we indicate at this stage by a subscript $w$ on the integration variable. When we
denote the current density in the interior $D_\alpha$ of object
$\alpha$ by $\bJ_\alpha$, we get
\begin{widetext}
\begin{eqnarray}
  \label{eq:Z-func-inte-sources}
  \cZ(\kappa)&=& \cZ_0 \prod_{\alpha=1}^N \int   [\cD \bJ_\alpha]_w [\cD \bJ^*_\alpha]_w 
\exp\left[-\frac{1}{2}\sum_{\alpha\beta} \int_{D_\alpha} \!\!d\bfx \int_{D_\beta} \!\!d\bfx'
\left\{ \bJ_\alpha^*(\bfx) \cG_>(\bfx,\bfx',\kappa) \bJ_\beta(\bfx') +\cc\right\}\right] 
\nonumber\\
&=& \cZ_0 \prod_{\alpha=1}^N \int   [\cD \bJ_\alpha]_w  [\cD \bJ^*_\alpha]_w\, e^{-\tilde S[\bJ]} \, ,
\end{eqnarray}
where we have used the continuity equation $\nabla\bJ=-i\kappa \rho$
to eliminate the charge density by introducing the tensor Green's
function for the half space
\begin{equation}
  \label{eq:Green-tensor}
  \cG_>(\bfx,\bfx',\kappa)= G_0(\bfx,\bfx',\kappa) \bI- G_0(\bfx,\bfx_R',\kappa) 
\begin{pmatrix}1&0&0\\0&1&0\\0&0&-1\end{pmatrix}
+\frac{1}{\kappa^2} \nabla\otimes \nabla' 
\left[G_0(\bfx,\bfx',\kappa)- G_0(\bfx,\bfx_R',\kappa)\right] \, ,
\end{equation}
\end{widetext}
where $G_0(\bfx,\bfx',\kappa)$ is the free scalar Green's function of
Eq.~\eqref{eq:G_free}, $\bI$ is the identity matrix and $\bfx_R$ is
defined below Eq.~\eqref{eq:free_G_D}.

The action of Eq.~\eqref{eq:Z-func-inte-sources} can be expressed
in terms of the original current densities $\bJ_\alpha$ and the mirror
current densities. Their components parallel and perpendicular to the
mirror plane are
\begin{equation}
  \label{eq:mirror-currents}
  \bJ_{\alpha,\|}^R(\bfx) = - \bJ_{\alpha,\|}(\bfx_R), \quad 
J_{\alpha,z}^R(\bfx) = J_{\alpha,z}(\bfx_R) \, .  
\end{equation}
The action then reads
\begin{eqnarray}
  \label{eq:Z-of-mirror-currents}
  \tilde S[\bJ] &=& \frac{1}{2} \sum_{\alpha\beta} \left\{ 
\int_{D_\alpha} d\bfx\int_{D_\beta} d\bfx' \bJ^*_\alpha(\bfx) \cG_0(\bfx,\bfx',\kappa)
\bJ_\beta(\bfx') \right. \nonumber\\
&+& \left.  
\int_{D_\alpha} \!\!d\bfx\int_{D^R_\beta} \!\!d\bfx' \bJ^*_\alpha(\bfx) \cG_0(\bfx,\bfx',\kappa)
\bJ^R_\beta(\bfx') +\cc\right\} \, .\nonumber\\
\end{eqnarray}
Now the sources are coupled by the free, infinite space Green's function 
\begin{equation}
\cG_0(\bfx,\bfx',\kappa) = G_0(\bfx,\bfx',\kappa) + \frac{1}{\kappa^2} 
\nabla\otimes \nabla' G_0(\bfx,\bfx',\kappa) \, .
\end{equation}
The mirror currents are located on the mirror objects $D_\alpha^R$
that are obtained from the original objects by reflection at the
mirror plane. This action has the same structure as in the case of
scalar fields. It can be expressed in terms of multipole moments of
the current densities very similarly to the scalar case. Again, we
consider diagonal and off-diagonal terms separately. 

{\it Off-diagonal terms} --- We introduce again local coordinates
$\bfx_\alpha$ that are measured relative to an origin inside
$D_\alpha$.  The terms that couple the original sources on different objects
can be written as
\begin{equation}
  \label{eq:EM-off-dia-terms}
  \tilde S_{\alpha\beta} = \frac{1}{2i\kappa} \int_{D_\alpha} d\bfx_\alpha 
\bJ^*_\alpha(\bfx_\alpha) \bE_\beta(\bfx_\alpha) + \cc 
\end{equation}
for $\alpha\neq\beta$ and the terms involving mirror sources become
\begin{equation}
  \label{eq:EM-off-dia-terms-mirror}
  \tilde S^R_{\alpha\beta} = \frac{1}{2i\kappa} \int_{D_\alpha} d\bfx_\alpha
\bJ^*_\alpha(\bfx_\alpha) \bE^R_\beta(\bfx_\alpha) + \cc  
\end{equation}
for all $\alpha$, $\beta$. Here we have introduced the electric fields
\begin{eqnarray}
  \label{eq:E-fields-of-sources}
  \bE_\beta(\bfx) &=& i\kappa \int_{D_\beta} d\bfx' \cG_0(\bfx,\bfx',\kappa) \bJ_\beta(\bfx')\\
  \bE^R_\beta(\bfx) &=& i\kappa \int_{D_\beta^R} d\bfx' \cG_0(\bfx,\bfx',\kappa) \bJ^R_\beta(\bfx')
\end{eqnarray}
which are generated by the current densities. For positions
$\bfx_\beta$ that are located outside a sphere that encloses the
object $D_\beta$, the electric field can be expressed in terms of the
electric and magnetic multipole moments of the current density
$\bJ_\beta$ for $l\ge 1$, $|m|\le l$,
\begin{eqnarray}
  \label{eq:mulitpole-moments-m}
  Q^\beta_{\textsc{m},lm}&\!=\!&  \int\!\! d \bfx_\beta \, 
\bJ_\beta(\bfx_{\alpha}) (-1)^l \bM_{lm}^{j*}(\bfx_\beta) \, ,\\
  \label{eq:mulitpole-moments-e}
   Q^\beta_{\textsc{e},lm}&\!=\!& \int\!\! d \bfx_\beta \, 
\bJ_\beta(\bfx_{\alpha}) (-1)^{l+1} \bN_{lm}^{j*}(\bfx_\beta) \, ,
\end{eqnarray}
where $\bM_{lm}^{j}(\bfx_\alpha)$, $\bN_{lm}^{j}(\bfx_\alpha)$
are the regular, divergence-less solutions
\begin{eqnarray}
  \label{eq:reg-vector-solutions}
\bM_{lm}^{j}(\bfx) &=& \frac{1}{\lambda}\nabla\times [\bfx j_l(i\kappa r)Y_{lm}(\hat\bfx)]
\\
\bN_{lm}^{j}(\bfx) &=& \frac{1}{\lambda}\frac{1}{i\kappa}
\nabla\times \nabla\times [\bfx j_l(i\kappa r)Y_{lm}(\hat\bfx)]
\end{eqnarray}
of the vector Helmholtz equation and $\lambda=\sqrt{l(l+1)}$. The electric
fields of the original sources can then be written as
\begin{equation}
  \label{eq:E-of-Mps-orig}
  \bE_\beta(\bfx_\beta) = -i\kappa^2 \sum_{lm} \left[
Q^\beta_{\textsc{m},lm} \bM_{lm}^h(\bfx_\beta)  + Q^\beta_{\textsc{e},lm} \bN_{lm}^h(\bfx_\beta)
\right] \, ,
\end{equation}
which is an expansion in the outgoing solutions
\begin{eqnarray}
  \label{eq:out-vector-solutions}
\bM_{lm}^{h}(\bfx) \!\!&=&\!\! \frac{1}{\lambda}\nabla\times [\bfx h_l^{(1)}(i\kappa r)Y_{lm}(\hat\bfx)]
\\
\bN_{lm}^{h}(\bfx) \!\!&=& \!\!\frac{1}{\lambda}\frac{1}{i\kappa}
\nabla\times \nabla\times [\bfx h_l^{(1)}(i\kappa r)Y_{lm}(\hat\bfx)]
\end{eqnarray}
of the vector Helmholtz equation. The electric fields of the mirror sources
can be also expressed in terms of the multipole moments of the original sources,
\begin{equation}
  \label{eq:E-of-Mps-mirror}
  \bE^R_\beta(\bfx_\beta^R) = -i\kappa^2 \sum_{lm} \left[
Q^{\beta,R}_{\textsc{m},lm} \bM_{lm}^h(\bfx^R_\beta)  + Q^{\beta,R}_{\textsc{e},lm} 
\bN_{lm}^h(\bfx^R_\beta)
\right] \, ,
\end{equation}
where $\bfx_\beta^R$ denotes the local coordinates of the mirror
object $D_\beta^R$.  Using Eq.~\eqref{eq:mirror-currents}, the
definition of the vector solutions of
Eq.~\eqref{eq:reg-vector-solutions} and $Y_{lm}(\hat\bfx_R)=(-1)^{l+m}
Y_{lm}(\hat\bfx)$, the moments of the mirror currents are given by
\begin{eqnarray}
  \label{eq:moments_of_mirrors-m}
  Q^{\beta,R}_{\textsc{m},lm} &=& (-1)^{l+m} Q^{\beta}_{\textsc{m},lm}\\
  \label{eq:moments_of_mirrors-e}
  Q^{\beta,R}_{\textsc{e},lm} &=& -(-1)^{l+m} Q^{\beta}_{\textsc{e},lm} \, .
\end{eqnarray}
The expansion of the electric fields into outgoing vector waves with
respect to the origin of the object that generates the field does not
allow us to perform the integrations in
Eqs.~\eqref{eq:EM-off-dia-terms}, \eqref{eq:EM-off-dia-terms-mirror}.
We would like to expand the electric field generated by object
$D_\beta$ in terms of vector waves that are regular at the origin of object
$D_\alpha$. When the coordinate systems associated with the two
objects have identical orientation, this can be done by relating the
outgoing and regular vector waves by a translation matrix ${\mathbb
  U}$.  Generalizing the result for scalar fields, the translation
matrix couples both types of vector solutions,
\begin{widetext}
\begin{eqnarray}
  \label{eq:EM-translation-matrix-def-m}
  \bM^h_{lm}(\bfx_\beta) &=& \sum_{l'm'} \left\{ B_{l'm'lm}(\bfX_{\alpha\beta}) \bM^j_{l'm'}
(\bfx_\alpha) + C_{l'm'lm}(\bfX_{\alpha\beta}) \bN^j_{l'm'} (\bfx_\alpha) \right\}\\
  \label{eq:EM-translation-matrix-def-n}
  \bN^h_{lm}(\bfx_\beta) &=& \sum_{l'm'} \left\{ B_{l'm'lm}(\bfX_{\alpha\beta}) \bN^j_{l'm'}
(\bfx_\alpha) + C_{l'm'lm}(\bfX_{\alpha\beta}) \bM^j_{l'm'} (\bfx_\alpha) \right\} 
\end{eqnarray}
with $\bfX_{\alpha\beta}=\bfx_\alpha-\bfx_\beta$.  The matrix elements
are given by \cite{Wittmann88},
\begin{eqnarray}
  \label{eq:EM-trans-mat-elements-gen-B}
  B_{l'm'lm}(\bfX_{\alpha\beta}) &=&  (-1)^{m} i^{l-l'} \sum_{l''} \frac{i^{l''}}{2}
\left[l(l+1)+l'(l'+1)-l''(l''+1)\right] \sqrt{\frac{(2l+1)(2l'+1)(2l''+1)}{l(l+1)l'(l'+1)}}
\nonumber\\
&&\times\begin{pmatrix}l&l'&l''\\0&0&0\end{pmatrix}
\begin{pmatrix}l&l'&l''\\m&-m'&m'-m\end{pmatrix}
h^{(1)}_{l''}(i\kappa|\bfX_{\alpha\beta}|) Y_{l''m-m'}(\hat\bfX_{\alpha\beta}) \, ,\\
  \label{eq:EM-trans-mat-elements-gen-C}
C_{l'm'lm}(\bfX_{\alpha\beta}) &=& 
\frac{\kappa}{\sqrt{l(l+1)l'(l'+1)}} \bfX_{\alpha\beta} \cdot \bigg[
\hat\bfx \frac{1}{2} \left(\lambda^+_{lm} A_{l'm'lm+1}(\bfX_{\alpha\beta}) +
\lambda^-_{lm} A_{l'm'lm-1}(\bfX_{\alpha\beta}) \right)\nonumber\\
&& \hspace*{3.6cm}+\, \hat\bfy \frac{1}{2i} \left(\lambda^+_{lm} 
A_{l'm'lm+1}(\bfX_{\alpha\beta}) -
\lambda^-_{lm} A_{l'm'lm-1}(\bfX_{\alpha\beta}) \right)\nonumber\\
&& \hspace*{3.6cm}+\, \hat\bfz\, m \,A_{l'm'lm}(\bfX_{\alpha\beta})
\bigg]
\end{eqnarray}
with
\begin{eqnarray}
  \label{eq:A-mtrix-elements}
A_{l'm'lm}(\bfX_{\alpha\beta}) &=& (-1)^m i^{l-l'} \sum_{l''} i^{l''} 
\sqrt{(2l+1)(2l'+1)(2l''+1)} \nonumber\\
&&\times \begin{pmatrix}l&l'&l''\\0&0&0\end{pmatrix}
\begin{pmatrix}l&l'&l''\\m&-m'&m'-m\end{pmatrix}
h^{(1)}_{l''}(i\kappa|\bfX_{\alpha\beta}|) Y_{l''m-m'}(\hat\bfX_{\alpha\beta})  
\end{eqnarray}
\end{widetext}
and $\lambda^\pm_{lm}=\sqrt{(l\mp m)(l\pm m+1)}$. It is useful to 
combine the translation matrix elements to the matrix
\begin{equation}
  \label{eq:def-EM-U-matrix}
  \cU^{\alpha\beta}_{l'm'lm} \equiv \begin{pmatrix}
B_{l'm'lm} & C_{l'm'lm}\\
-C_{l'm'lm} & -B_{l'm'lm}\\
\end{pmatrix}(\bfX_{\alpha\beta})
\end{equation}
that acts on the vector
$\bQ^\alpha_{lm}=(Q^\alpha_{\textsc{m},lm},Q^\alpha_{\textsc{e},lm})$
of multipole moments.

Using the translation formula of
Eqs.~\eqref{eq:EM-translation-matrix-def-m},
\eqref{eq:EM-translation-matrix-def-n}, we can express the field
$\bE_\beta(\bfx_\alpha)$ in Eq.~\eqref{eq:EM-off-dia-terms} in terms
of regular vector solutions and perform the integration. This
integration yields the multipole moments defined by
Eqs.~\eqref{eq:mulitpole-moments-e}, \eqref{eq:mulitpole-moments-m}.
The action of Eq.~\eqref{eq:EM-off-dia-terms} can then be written as
\begin{equation}
  \label{eq:EM-off-dia-mps}
\sum_{\alpha\neq\beta} \tilde S_{\alpha\beta} = -\kappa \sum_{\alpha\neq\beta} \sum_{lml'm'} 
\bQ^{\alpha *}_{l'm'} \tilde \cU^{\alpha\beta}_{l'm'lm} \bQ^\beta_{lm}  \, ,  
\end{equation}
where we have defined the modified translation matrix
\begin{equation}
  \label{eq:EM-U-mat-mod}
  \tilde\cU^{\alpha\beta}_{l'm'lm} \equiv \frac{1}{2}\left[
(-1)^{l'}\cU^{\alpha\beta}_{l'm'lm} + (-1)^l\cU^{{\beta\alpha}^\dagger}_{lml'm'} \right] \, ,
\end{equation}
where $\dagger$ denotes the conjugate transpose of the matrix of
Eq.~\eqref{eq:def-EM-U-matrix}. For translations along the z-axis,
$\bfX_{\alpha\beta} \sim \hat \bfz$, the latter expression is
diagonal in $m$ and simplifies to $\tilde\cU^{\alpha\beta}_{l'mlm} =
(-1)^{l'}\cU^{\alpha\beta}_{l'mlm}$ due to the symmetries
$B^*_{lml'm}(\bfX_{\beta\alpha})= (-1)^{l+l'}
B_{l'mlm}(\bfX_{\alpha\beta})$ and $C^*_{lml'm}(\bfX_{\beta\alpha})= -(-1)^{l+l'}
C_{l'mlm}(\bfX_{\alpha\beta})$. 

The action of Eq.~\eqref{eq:EM-off-dia-terms-mirror} that couples the
original and mirror sources can be cast into a form similar to
Eq.~\eqref{eq:EM-off-dia-mps} with a translation matrix ${\mathbb
  U}^R$ that is defined by
\begin{equation}
  \label{eq:def-EM-U-matrix-mirror}
  \cU^{R,\alpha\beta}_{l'm'lm} = (-1)^{l+m}\begin{pmatrix}
B_{l'm'lm} & -C_{l'm'lm}\\
-C_{l'm'lm} & B_{l'm'lm}\\
\end{pmatrix}(\bfx_\alpha-\bfx_\beta^R) \, ,
\end{equation}
where the phase factors relating the original and mirror multipole
moments in Eqs.~\eqref{eq:moments_of_mirrors-m},
\eqref{eq:moments_of_mirrors-e}  have been absorbed in the translation matrix.
With this definition Eq.~\eqref{eq:E-of-Mps-mirror} becomes
\begin{equation}
  \label{eq:EM-off-diag-terms-mirror-mps}
\sum_{\alpha\beta}  \tilde S_{\alpha\beta}^R =  \kappa \sum_{\alpha\beta} \sum_{lml'm'} 
\bQ^{\alpha *}_{l'm'} \tilde \cU^{R,\alpha\beta}_{l'm'lm} \bQ^\beta_{lm} 
\end{equation}
with  the modified translation matrix 
\begin{equation}
  \label{eq:EM-U-mat-mod-mirror}
  \tilde\cU^{R,\alpha\beta}_{l'm'lm} \equiv \frac{1}{2}\left[
(-1)^{l'}\cU^{R,\alpha\beta}_{l'm'lm} + (-1)^l\cU^{{R,\beta\alpha}^\dagger}_{lml'm'} \right] \, .
\end{equation}
For $\alpha=\beta$, the action of
Eq.~\eqref{eq:EM-off-diag-terms-mirror-mps} describes the interaction
between a source and its mirror image. In this case the translation
vector is $\bfx_\alpha-\bfx_\alpha^R = -2L_\alpha \hat\bfz$ where
$L_\alpha$ is the normal distance between the origin of object
$D_\alpha$ and the mirror plane.  Hence, the translation is along the
z-axis and the translation matrix elements of
Eqs.~\eqref{eq:EM-trans-mat-elements-gen-B},
\eqref{eq:EM-trans-mat-elements-gen-C} simplify to
\begin{widetext}
\begin{eqnarray}
  \label{eq:EM-trans-mat-elements-z-B}
  B_{l'm'lm}(-2L_\alpha\hat\bfz) &=& \delta_{m'm} (-1)^{m} i^{l-l'} \sum_{l''} \frac{i^{-l''}}{2} 
\left[l(l+1)+l'(l'+1)-l''(l''+1)\right] \sqrt{\frac{(2l+1)(2l'+1)}{l(l+1)l'(l'+1)}}
(2l''+1)\nonumber\\
&&\times\begin{pmatrix}l&l'&l''\\0&0&0\end{pmatrix}
\begin{pmatrix}l&l'&l''\\m&-m&0\end{pmatrix}
h^{(1)}_{l''}(i\kappa 2L_\alpha)  \, ,\\
  \label{eq:EM-trans-mat-elements-z-C}
C_{l'm'lm}(-2L_\alpha\hat\bfz) &=& -\delta_{m'm} \kappa 2L_\alpha m (-1)^m 
i^{l-l'} \sum_{l''} i^{-l''} 
\sqrt{\frac{(2l+1)(2l'+1)}{l(l+1)l'(l'+1)}}
(2l''+1)\nonumber\\
&&\times\begin{pmatrix}l&l'&l''\\0&0&0\end{pmatrix}
\begin{pmatrix}l&l'&l''\\m&-m&0\end{pmatrix}
h^{(1)}_{l''}(i\kappa 2L_\alpha) \, .
\end{eqnarray}
\end{widetext}
These matrix elements obey the symmetry relations
$B^*_{l'mlm}(-2L_\alpha\hat\bfz)=B_{lml'm}(-2L_\alpha\hat\bfz)$,
$C^*_{l'mlm}(-2L_\alpha\hat\bfz)=C_{lml'm}(-2L_\alpha\hat\bfz)$ so
that the matrix of Eq.~\eqref{eq:EM-U-mat-mod-mirror}
simplifies for $\alpha=\beta$ to $\tilde \cU^{R,\alpha\alpha}_{l'mlm}=
(-1)^{l'} \cU^{R,\alpha\alpha}_{l'mlm}$. 

{\it Diagonal terms} --- So far we have described the interaction
between {\it arbitrary} multipoles inside the material objects. But
the functional integral over currents of
Eq.~\eqref{eq:Z-func-inte-sources} contains weights that measure the
energy cost for inducing currents on an object with particular shape
and material composition.  In the following we will show that these
weights can be implemented straightforwardly when we express the
diagonal terms of the action of Eq.~\eqref{eq:Z-of-mirror-currents},
\begin{equation}
  \label{eq:EM-diagonal-action}
  \tilde S_\alpha = \frac{1}{2} \int_{D_\alpha} \!\!d\bfx \int_{D_\alpha} \!\!d\bfx' \left\{
\bJ^*_\alpha(\bfx) \cG_0(\bfx,\bfx',\kappa)\bJ_\alpha(\bfx') + \cc \right\} \, ,
\end{equation}
in terms of the incident field which generates a prescribed current
that corresponds to the induced polarization and magnetization inside
the object. Then the relation between the induced current and the
incident field ensures that the currents are weighted properly. To see
this we employ the macroscopic formulation of the electromagnetic
action, see Eq.~\eqref{eq:EM-S-macro}. The energy of
Eq.~\eqref{eq:EM-diagonal-action} is associated with the process of
building up the induced current $\bJ_\alpha$ inside the object. This
energy must therefore be equal to the change in the total macroscopic
field energy that results when the object is placed into an external
field.  Thus we obtain
\begin{equation}
  \label{eq:EM-diagonal-action-macroscopic}
  \tilde S_\alpha = \frac{1}{2} \int \!d\bfx \left[
\bE \bD^* +\bB \bH^* - \left( \bE_0 \bD_0^* +\bB_0 \bH_0^* \right) + \cc
\right] \, ,
\end{equation}
where integration extends over all space. Field vectors with
subscript $0$ represent the incident field that is generated by some
fixed external sources in otherwise empty space and field vectors
without label stand for the total field from the external and induced
sources after adding the object. $\tilde S_\alpha$ can be also written
as
\begin{eqnarray}
  \label{eq:EM-mac-action}
\tilde S_\alpha &=& \frac{1}{2} \int \!d\bfx \left[
\bE \bD_0^* -\bD^* \bE_0 + \bB \bH_0^* - \bH^* \bB_0 + \cc
\right] \nonumber\\
&+&  \frac{1}{2} \int \!d\bfx \left[ (\bE+\bE_0) (\bD^*-\bD^*_0) \right. \nonumber\\
&& \quad \quad\,\,\,\left. + (\bB+\bB_0) ( \bH^* -\bH_0^*) +\cc \right] \, . 
\end{eqnarray}
The second integral of this expression vanishes. This can be seen by
setting $\bE + \bE_0 = i\kappa \bA - \nabla \Phi$, $\bB+\bB_0 =
\nabla \times \bA$ and using that $\nabla\bD^* = \rho^*_\text{ext} =
\nabla\bD_0^*$, $\nabla\times \bH^* +i\kappa \bD^* = \bJ^*_\text{ext}=
\nabla\times \bH_0^* +i\kappa \bD_0^*$ since the external sources
$\rho_\text{ext}$, $\bJ_\text{ext}$ are fixed when the object is
added. Notice that the bound (induced) sources do not appear
explicitly since they are included in the macroscopic fields.  Since
$\bD_0=\bE_0$, $\bH_0=\bB_0$ and $\bD=\epsilon \bE$, $\bH=\mu^{-1}
\bB$ with $\epsilon=\epsilon_\alpha$, $\mu=\mu_\alpha$ inside the
objects and $\epsilon=1$, $\mu=1$ outside the objects, we get
\begin{equation}
  \label{eq:EM-mac-action-2}
  \tilde S_\alpha = \frac{1}{2} \int_{D_\alpha} \!d\bfx \left[
(1-\epsilon_\alpha) \bE \bE_0^* + (1-\mu_\alpha^{-1})\bB \bB_0^* + \cc
\right] \, ,
\end{equation}
where integration runs only over the interior of the object.  The
material dependent functions $\epsilon_\alpha$ and $\mu_\alpha$ can be
expressed in terms of the polarization $\bP_\alpha$ and magnetization
$\bM_\alpha$ of the object. The relations between macroscopic fields
yield $\bP_\alpha = \bD-\bE = (\epsilon_\alpha-1)\bE$ and
$\bM_\alpha=\bB - \bH = (1-1/\mu_\alpha) \bB$. When we substitute
$\bB_0^* = i/\kappa \nabla\times \bE_0^*$ in Eq.~(\ref{eq:EM-mac-action-2}) we
can integrate by parts to obtain
\begin{equation}
  \label{eq:EM-mac-action-final}
  \tilde S_\alpha = -\frac{1}{2} \int_{D_\alpha} \!d\bfx \left[
\frac{1}{i\kappa} \bJ_\alpha(\bfx) \bE_0^*(\bfx) + \cc 
\right] \, ,
\end{equation}
where we have combined the polarization and magnetization to
yield the induced current density
\begin{equation}
  \label{eq:ind_J_from_P+M}
  \bJ_\alpha = i\kappa \bP_\alpha + \nabla \times \bM_\alpha \, .
\end{equation}
Notice that the incident field $\bE_0$ in
Eq.~(\ref{eq:EM-mac-action-final}) depends on the current density
$\bJ_\alpha$ since $\bE_0$ has to induce the prescribed current
density. Hence, the problem of expressing the diagonal part of the
action in terms of multipole moments has been reduced to computing the
incident field that has to act on the object to generate a given
current density. In scattering theory one usually encounters the
opposite problem. For an incident field one would like to compute the
scattered field which can be expanded in outgoing partial vector
waves, see Eq.~(\ref{eq:E-of-Mps-orig}). Here the situation is
slightly different. We seek to determine the incident field that
generates a given set of multipole moments inside the object.  In
other words, for a given scattered field, which according to
Eq.~(\ref{eq:E-of-Mps-orig}) is given by the multipole moments, we
would like to obtain the corresponding incident field.  We expand
the incident field as
\begin{equation}
  \label{eq:E-inci-expanded}
  \bE_0(\bfx) = -i\kappa^2 \sum_{lm} \left[
\phi_{\textsc{m},lm} \bM_{lm}^j(\bfx)  + \phi_{\textsc{e},lm} \bN_{lm}^j(\bfx)
\right] \, .
\end{equation}
The relation between the multipole moments and the amplitudes of the
incident field is determined by the T-matrix ${\mathbb T}^\alpha=
({\mathbb S}^\alpha -1)/2$, where ${\mathbb S}^\alpha$ is the
scattering matrix of the object. When we solve this relation for the
incident field amplitudes
$\bphi_{lm}=(\phi_{\textsc{m},lm},\phi_{\textsc{e},lm})$, we get
\begin{equation}
  \label{eq:invs-T-matrix-relation}
  \bphi_{lm} = \sum_{l'm'} [\cT^\alpha]^{-1}_{lml'm'} \bQ^\alpha_{l'm'} \, ,
\end{equation}
where $\cT^\alpha_{lml'm'}$ is a $2\times 2$ matrix acting on magnetic
and electric multipoles.  This relation together with
Eq.~(\ref{eq:E-inci-expanded}) yields the incident field that generates the
multipoles $\bQ^\alpha_{lm}$. With this result we can perform the
integration in Eq.~(\ref{eq:EM-mac-action-final}) which yields 
$\tilde S_\alpha$ in terms of the multipoles,
\begin{equation}
\label{eq:EM-diag-action-of-MPs}
\tilde S_\alpha = \kappa \!\!\sum_{lml'm'}\!\! \bQ^{\alpha *}_{lm}\,  
\frac{1}{2} \left\{ [\tilde \cT^\alpha ]^{-1}_{lml'm'} +  
\left[ [\tilde \cT^\alpha ]^{-1}\right]^\dagger_{l'm'lm}  \right\}
 \bQ^{\alpha}_{l'm'}  
\end{equation}
with the modified T-matrix
\begin{equation}
  \label{eq:EM-mod-T}
  \tilde \cT^\alpha_{lml'm'} =  (-1)^{l'}
\begin{pmatrix}
-T^{\alpha,\textsc{mm}}_{lml'm'}&T^{\alpha,\textsc{me}}_{lml'm'}\\[1em]
-T^{\alpha,\textsc{em}}_{lml'm'}&T^{\alpha,\textsc{ee}}_{lml'm'}
\end{pmatrix} \, ,
\end{equation}
where the $T^{\alpha,XY}_{lml'm'}$ are the elements of the matrix
$\cT^\alpha_{lml'm'}$ with $X$, $Y=\textsc{m}$, $\textsc{e}$ labeling
magnetic and electric elements, respectively. Due to the symmetry
\begin{equation}
  \label{eq:EM-T-symmetry}
  \begin{pmatrix}
T^{\alpha,\textsc{mm}}_{lml'm'}&T^{\alpha,\textsc{me}}_{lml'm'}\\[1em]
T^{\alpha,\textsc{em}}_{lml'm'}&T^{\alpha,\textsc{ee}}_{lml'm'}
\end{pmatrix}^* \!\!=(-1)^{l+l'}
\begin{pmatrix}
T^{\alpha,\textsc{mm}}_{l'm'lm}&-T^{\alpha,\textsc{em}}_{l'm'lm}\\[1em]
-T^{\alpha,\textsc{me}}_{l'm'lm}&T^{\alpha,\textsc{ee}}_{l'm'lm}
\end{pmatrix}
\end{equation}
of the T-matrix Eq.~(\ref{eq:EM-diag-action-of-MPs}) can be
simplified to
\begin{equation}
\label{eq:EM-diag-action-of-MPs-simp}
\tilde S_\alpha = \kappa \!\!\sum_{lml'm'}\!\! \bQ^{\alpha *}_{lm}\,  
 [\tilde \cT^\alpha ]^{-1}_{lml'm'}   
 \bQ^{\alpha}_{l'm'}  \, .
\end{equation}

The total action $\tilde S[\bQ]$ for the multipole moments is given by
the sum of the actions of Eqs.~(\ref{eq:EM-off-dia-mps}),
(\ref{eq:EM-off-diag-terms-mirror-mps}) and
(\ref{eq:EM-diag-action-of-MPs-simp}). In compact matrix notation it can be written
as 
\begin{equation}
  \label{eq:EM-total_action-final}
  \tilde S[\bQ] = \sum_{\alpha\beta} \bQ^*_\alpha {\mathbb M}^{\alpha\beta} \bQ_\beta \, ,
\end{equation}
with the matrix
\begin{equation}
  \label{eq:EM-def-matrix-M}
  {\mathbb M}^{\alpha\beta}= \kappa \left\{ [\tilde {\mathbb T}^\alpha ]^{-1} 
\delta_{\alpha\beta} - \tilde {\mathbb U}^{\alpha\beta} (1-\delta_{\alpha\beta})
+ \tilde {\mathbb U}^{R,\alpha\beta} \right\} \, ,
\end{equation}
where $\tilde {\mathbb T}^\alpha$, $\tilde {\mathbb U}^{\alpha\beta}$
and $\tilde {\mathbb U}^{R,\alpha\beta}$ stand for the matrices with
matrix elements given by Eqs.~(\ref{eq:EM-mod-T}),
(\ref{eq:EM-U-mat-mod}) and (\ref{eq:EM-U-mat-mod-mirror}),
respectively. In analogy to the scalar case,
Eq.~(\ref{eq:Z-og-kappa-final}), the partition function $\cZ(\kappa)$
is obtained by integrating over all multipoles. Notice that by construction
of the diagonal parts of the action of Eq.~(\ref{eq:EM-total_action-final}) 
the proper weights are assigned to the multipole configurations. 

The Gaussian integral over multipoles and Eq.~(\ref{eq:casimir-energy})
lead to the final result 
\begin{equation}
  \label{eq:EM-Casimir-energy-gen-result}
  \cE = \frac{\hbar c}{2\pi} \int_0^\infty d\kappa \ln\frac{\det {\mathbb M}}
{\det {\mathbb M}_\infty} 
\end{equation}
for the electromagnetic Casimir energy where the determinant is taken
with respect to the partial wave indices $l$, $m$, polarization
indices $\textsc{m}$, $\textsc{e}$ and the object indices $\alpha$, $\beta$.
The matrix ${\mathbb M}_\infty$ is given by Eq.~(\ref{eq:EM-def-matrix-M}) with
the translation matrices $\tilde {\mathbb U}^{\alpha\beta}$
and $\tilde {\mathbb U}^{R,\alpha\beta}$ set to zero.

For one object in front of a perfectly reflecting mirror plane, the
Casimir energy can be written as
\begin{equation}
  \label{eq:EM-casimir-energy-1-surface}
  \cE_1 = \frac{\hbar c}{2\pi} \int_0^\infty d\kappa \ln \det \left(
1 + \tilde {\mathbb T}^1 \tilde {\mathbb U}^{R,11} \right) \, .
\end{equation}
The matrix $\tilde {\mathbb U}^{R,11}$ that describes the
interaction between fluctuating currents on the object and its mirror
image is universal, and the shape and material composition of the object
enters through its T-matrix $\tilde {\mathbb T}^1$.

\section{Plate -- sphere geometry}
\label{sec:plate-sphere}

In this Section we consider a geometry that is most relevant to a
large number of experimental studies of Casimir interactions carried
out in the last decade. This geometry, consisting of a plane mirror
and a sphere, is experimentally favorable since it avoids the problem
of parallelism for plane surfaces facing each other. Despite its
experimental importance, this geometry lacks a theoretical description
of the electromagnetic Casimir interaction. For a scalar field with
Dirichlet boundary conditions at the sphere and the plane, the
interaction over a wide range of distances has been obtained recently
\cite{Bulgac+06,Wirzba:qfext07}, including an analytic expression for
the lowest order correction to the proximity force approximation
\cite{Bordag:qfext07}.

\begin{figure}
\includegraphics[scale=0.7]{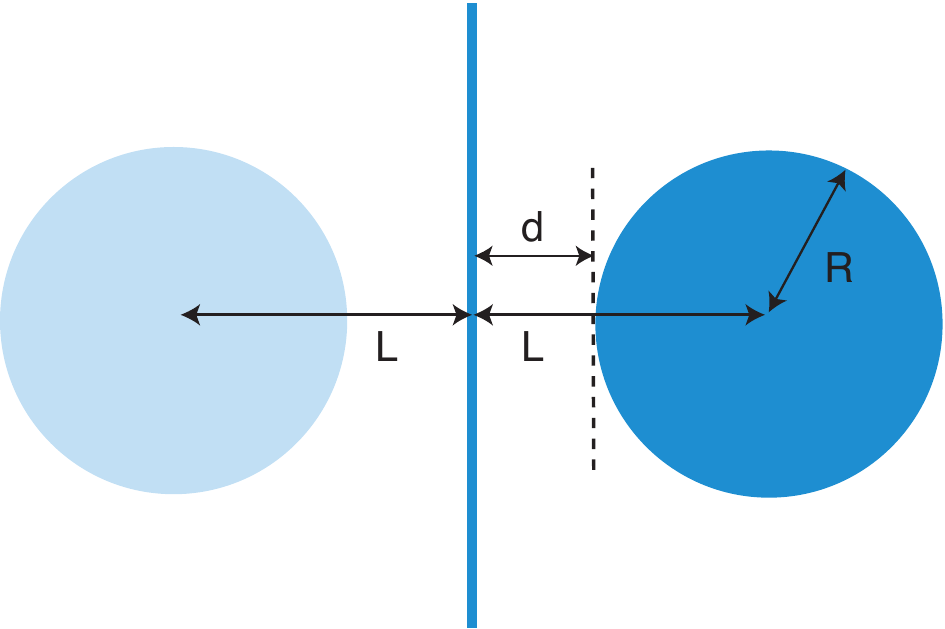}
\caption{Length scales of the sphere-mirror geometry.}
\label{fig:plane-sphere}
\end{figure}

We consider a sphere of radius $R$ in front of a plane mirror with
distance $L$ between the center of the sphere and the mirror, see
Fig.~\ref{fig:plane-sphere}. For the scalar field we study Dirichlet
or Neumann boundary conditions at the mirror and the sphere. For the
electromagnetic field, the sphere consists of a material with
arbitrary dielectric function $\epsilon(k)$ and permeability $\mu(k)$
while the mirror is assumed to be perfectly reflecting (implying
vanishing of the parallel components of the electric field and the
normal component of the magnetic field).

\subsection{Large distance expansion}

We start by studying the Casimir interaction at large separations.
For separations $L$ that are large compared to the size $R$ of the object,
the Casimir energy can be expressed as an asymptotic series in $R/L$.
Using that $\ln\det = \tr \ln$ in Eqs.~(\ref{eq:casimir-energy-1-surface}) and
(\ref{eq:EM-casimir-energy-1-surface}), we get by expanding the logarithm
\begin{eqnarray}
  \label{eq:energy-scatt-exp}
  \cE &=& \frac{\hbar c}{2\pi} \int_0^\infty d\kappa 
\tr \ln ( 1 - {\mathbb N}) \\
&=& - \frac{\hbar c}{2\pi} \int_0^\infty d\kappa \sum_{p=1}^\infty 
\frac{1}{p} \tr\left(\mathbb{N}^p\right) \, ,
\end{eqnarray}
where we have defined the operator $\mathbb{N}= \mp\mathbb{T}^{1}
\mathbb{U}^{R,11}$ with $-$($+$) for a scalar field in the presence of
a Dirichlet (Neumann) mirror and $\mathbb{N}= -\tilde {\mathbb T}^{1}
\tilde {\mathbb U}^{R,11}$ for the electromagnetic field and a perfectly
reflecting mirror. The operator ${\mathbb N}$ describes a wave that travels
from the mirror to the object and back, involving one scattering at the 
object.

\subsubsection{Scalar field}

All information about the shape of the surface and the boundary conditions
is provided by the ${\mathbb T}$-matrix. For a spherically symmetric 
surface the matrix is diagonal and completely specified by scattering
phase shifts $\delta_l(k)$ that do not depend on $m$,
\begin{equation}
  \label{eq:spheres-t-from-shifts}
  \cT_{lml'm'}(k)=\delta_{ll'}\delta_{mm'} \frac{1}{2} 
\left( e^{2i\delta_l(k)}-1\right) \, , 
\end{equation}
where $k$ is the real frequency.  The scattering phase shifts for a
sphere of radius $R$ with Dirichlet and Neumann boundary conditions
are
\begin{eqnarray}
  \label{eq:sphere-phases-D}
  \cot \delta^D_l(k) &=& \frac{n_l(\xi)}{j_l(\xi)} \, ,\\
  \label{eq:sphere-phases-N}
  \cot \delta^N_l(k) &=& \frac{n'_l(\xi)}{j'_l(\xi)} \, ,
\end{eqnarray}
where $\xi=kR$ and $j_l$ ($n_l$) are spherical Bessel functions of
first (second) kind. To make use of the general result of
Eq.~\eqref{eq:casimir-energy-1-surface}, we have to evaluate the
matrix elements for imaginary frequencies $k=i\kappa$. This can be
done using $j_l(iz)=i^l \sqrt{\pi/(2z)} I_{l+1/2}(z)$,
$h_l^{(1)}(iz)=j_l(iz)+in_l(iz)=-i^{-l} \sqrt{2/(\pi z)}
K_{l+1/2}(z)$, where $I_{l+1/2}$ and $K_{l+1/2}$ are modified Bessel
functions of first and second kind, respectively. We obtain
\begin{eqnarray}
  \label{eq:T-matrix-sphere-of-kappa}
  \cT^D_{lmlm} &=& (-1)^l \frac{\pi}{2} \frac{I_{l+1/2}(z)}{K_{l+1/2}(z)}\\
  \cT^N_{lmlm} &=& (-1)^l \frac{\pi}{2} \frac{I_{l+1/2}(z)-2z I'_{l+1/2}(z)}
{K_{l+1/2}(z)-2z K'_{l+1/2}(z)}
\end{eqnarray}
for Dirichlet and Neumann boundary conditions where $z=\kappa R$.

The distance dependence of the Casimir energy of
Eq.~\eqref{eq:casimir-energy-1-surface} is given by the translation
matrix that describes the interaction between the original and mirror
multipole moments. Since the origins (centers) of the original and
mirror sphere are related by a translation along the $z$-axis, the
matrix elements of the translation matrix ${\mathbb U}^{R,11}$ are
given by Eqs.~\eqref{eq:U-matrix-elements-R} and
\eqref{eq:U-matrix-elements-z-direction}. After a rotation of
frequency to the imaginary axis we obtain
\begin{widetext}
\begin{equation}
  \label{eq:transl-matrix-imag}
\cU^{R,11}_{l'mlm}= 
- (-1)^{l} i^{-l'+l} \sqrt{(2l+1)(2l'+1)}\, \sum_{l''} \,
 (-1)^{l''} (2l''+1)
\begin{pmatrix}l&l'&l''\\0&0&0\end{pmatrix}
\begin{pmatrix}l&l'&l''\\m&-m&0\end{pmatrix}
\frac{K_{l''+1/2}(2\kappa L)}{\sqrt{\pi\kappa L}} \, .
\end{equation}
\end{widetext}
With the elements of the transition and translation matrices we can
compute the Casimir interaction from
Eq.~\eqref{eq:casimir-energy-1-surface}. The scaling of the ${\mathbb
  T}$-matrix for small $\kappa$ shows that partial waves of order $l$
start to contribute to the energy at order $L^{-2(l+1)}$ if the
${\mathbb T}$-matrix is diagonal in $l$. Hence, we can truncate the
infinite matrix ${\mathbb N}$ in Eq.~\eqref{eq:energy-scatt-exp} at
finite multipole order to obtain the large distance expansion of the
energy at a given order in $R/L$. Also, we can truncate the series
over $p$ in Eq.~\eqref{eq:energy-scatt-exp} since the $p^\text{th}$
power of ${\mathbb N}$ becomes important only at order $L^{-(p+1)}$.
The first $p$ powers of ${\mathbb N}$ describe $p$ scattering off
the sphere. 

We have used {\tt Mathematica} to perform the matrix operations
and to expand the integrand of Eq.~\eqref{eq:energy-scatt-exp} 
in $R/L$. From this we find that the Casimir energy can be
written as
\begin{equation}
\label{eq:spheres-energy-large-L}
\cE = \frac{\hbar c }{\pi} \frac{1}{L}
\sum_{j=2}^\infty b_j \left(\frac{R}{L}\right)^{j-1} \, ,
\end{equation}
where $b_j$ is the coefficient of the term $\sim L^{-j}$. In the
following we give the results for a mirror plane with Dirichlet
or Neumann boundary conditions separately.

{\it Dirichlet mirror} --- In this case we have $\mathbb{N}=
-\mathbb{T}^{1} \mathbb{U}^{R,11}$. If the field obeys Dirichlet
boundary conditions at the sphere, we obtain the for leading
coefficients
\begin{eqnarray}
\label{eq:coeff_D-D}
b_2&=&-\frac{1}{8}, \quad
b_3=-\frac{5}{64}, \quad b_4=-\frac{421}{1152}, 
\nonumber\\ 
\quad b_5&=&-\frac{535}{9216},
\quad  b_6=-\frac{3083041}{4147200}, \quad 
b_7=+\frac{2741117}{11059200}, 
\nonumber\\ 
\quad b_8&=&-\frac{557222415727}{292626432000} \, ,
\end{eqnarray}
indicating an attractive force. This result agrees with recent
findings of Wirzba \cite{Wirzba:qfext07}.  If we impose Neumann
boundary conditions at the sphere, we find
\begin{eqnarray}
\label{eq:coeff_D-N}
b_2&=&0, \quad
b_3=0, \quad b_4=\frac{17}{96}, 
\nonumber\\ 
\quad b_5&=&0,
\quad  b_6=\frac{379}{960}, \quad 
b_7=\frac{479}{24576}, 
\nonumber\\ 
\quad b_8&=&\frac{243487}{403200} \, .
\end{eqnarray}
This result corresponds to a repulsive force which is expected for
unlike boundary conditions.  To leading order the energy scales as
$R^3/L^4$ since the first two coefficients vanish. This behavior can
be understood from the absence of low-frequency $s$-wave scattering
for a sphere with Neumann boundary conditions.

{\it Neumann mirror} --- In this case we have $\mathbb{N}=
+\mathbb{T}^{1} \mathbb{U}^{R,11}$. For a Dirichlet sphere
we obtain the coefficients
\begin{eqnarray}
\label{eq:coeff_N-D}
b_2&=&\frac{1}{8}, \quad
b_3=\frac{3}{64}, \quad b_4=\frac{403}{1152}, 
\nonumber\\ 
\quad b_5&=&\frac{73}{9216},
\quad  b_6=\frac{2981791}{4147200}, \quad 
b_7=-\frac{3636227}{11059200}, 
\nonumber\\ 
\quad b_8&=&\frac{550781720977}{292626432000} \, ,
\end{eqnarray}
and hence the force is again repulsive. The modulus of the
coefficients are smaller than those in Eq.~\eqref{eq:coeff_D-D} for a
Dirichlet mirror with the exception of the leading one, $b_2$ and the
one with inverted sign, $b_7$. Notice that the sum of the Casimir
energies for the Dirichlet and Neumann mirror, both opposite to a
Dirichlet sphere, is identical to the Casimir energy between two
Dirichlet spheres at a center-to-center distance $2L$
\cite{Bulgac+06}. The reason for this is that any field configuration
can be decomposed into symmetric and antisymmetric modes with respect
to the mirror plane which obey Neumann and Dirichlet boundary
conditions on that plane, respectively.  Since the energy between two
spheres scales as $R^2/L^3$ for asymptotically large separations, the
modulus of the coefficients $b_2$ in Eqs.~(\ref{eq:coeff_D-D}) and
(\ref{eq:coeff_N-D}) must be identical so that the contribution $\sim
R/L^2$ cancels. It is easily seen that the higher order coefficients
of Eqs.~(\ref{eq:coeff_D-D}) and (\ref{eq:coeff_N-D}) combine to the
correct coefficients of the large distance expansion for two Dirichlet
spheres \cite{Emig+07}.

For a Neumann sphere we get
\begin{eqnarray}
\label{eq:coeff_N-N}
b_2&=&0, \quad
b_3=0, \quad b_4=-\frac{17}{96}, 
\nonumber\\ 
\quad b_5&=&0,
\quad  b_6=-\frac{379}{960}, \quad 
b_7=-\frac{267}{8192}, 
\nonumber\\ 
\quad b_8&=&-\frac{243487}{403200} \, .
\end{eqnarray}
For the same reason as before, the first two coefficients
vanish. Notice that all given coefficients, with the exception of
$b_7$, are equal to minus the corresponding coefficients for a Neumann
sphere and a Dirichlet mirror in Eq.~(\ref{eq:coeff_D-N}).  This
result is consistent with the observation that the Casimir energy for
two Neumann spheres at a center-to-center distance $2L$ is given by
the sum of the energies of a Neumann sphere at distance $L$ from a
Dirichlet mirror and a Neumann mirror, respectively. Since two Neumann
spheres interact at large distance with an energy $\sim R^6/L^7$ with
the next-to-leading term $\sim R^8/L^9$, the sum of the corresponding
coefficients, excepting $b_7$, must vanish. The coefficients $b_7$ in
Eqs.~(\ref{eq:coeff_D-N}), (\ref{eq:coeff_N-N}) combine to the correct
value for two Neumann spheres that was obtained in Ref.~\cite{Emig+07}.

\subsubsection{Electromagnetic field}

We consider a dielectric sphere at a center-to-surface distance $L$
from a perfectly conducting mirror.  Due to spherical symmetry, the electric and
magnetic multipoles for all $l$, $m$ are decoupled so that
the T-matrix is diagonal \cite{Emig+07b},
\begin{widetext}
\begin{equation}
\label{eq:t-matrix-elem-sphere}
  T^{\textsc{mm}}_{lmlm}=(-1)^l \frac{\pi}{2} \frac{\eta I_{l+{1\over 2}}(z)
\left[I_{l+{1\over 2}}(nz)+2nzI'_{l+{1\over 2}}(nz)\right] - n I_{l+{1\over 2}}(nz)
\left[I_{l+{1\over 2}}(z)+2z I'_{l+{1\over 2}}(z)\right]}
{\eta K_{l+{1\over 2}}(z)
\left[I_{l+{1\over 2}}(nz)+2nzI'_{l+{1\over 2}}(nz)\right] - n I_{l+{1\over 2}}(nz)
\left[K_{l+{1\over 2}}(z)+2z K'_{l+{1\over 2}}(z)\right]} \, ,
\end{equation}
\end{widetext}
where the sphere radius is $R$, $z=\kappa R$,
$n=\sqrt{\epsilon(i\kappa)\mu(i\kappa)}$,
$\eta=\sqrt{\epsilon(i\kappa)/\mu(i\kappa)}$.
$T^{\textsc{ee}}_{lmlm}$ is obtained from
Eq.~(\ref{eq:t-matrix-elem-sphere}) by interchanging $\epsilon$ and
$\mu$. The limit of a perfectly conducting sphere is obtained by
taking $\epsilon(i\kappa)\to\infty$ at an arbitrarily fixed
$\mu(i\kappa)$ which can also vanish.  Then the matrix elements become
independent of $\mu$,
\begin{eqnarray}
  \label{eq:t-matrix-elem-cond-sphere-m}
   T^{\textsc{mm}}_{lmlm}&=& (-1)^l \frac{\pi}{2} \frac{I_{l+{1\over 2}}(z)}{K_{l+{1\over 2}}(z)}
\\
  \label{eq:t-matrix-elem-cond-sphere-e}
T^{\textsc{ee}}_{lmlm}&=& (-1)^l \frac{\pi}{2} \frac{I_{l+{1\over 2}}(z)+2zI'_{l+{1\over 2}}(z)}
{K_{l+{1\over 2}}(z)+2zK'_{l+{1\over 2}}(z)} \, .
\end{eqnarray}
For all partial waves, the {\it leading} low frequency contribution is
determined by the {\it static} electric multipole polarizability,
$\alpha^\textsc{e}_l = [(\epsilon-1)/(\epsilon+(l+1)/l)]R^{2l+1}$, and
the corresponding magnetic polarizability, $\alpha^\textsc{m}_l =
[(\mu-1)/(\mu+(l+1)/l)]R^{2l+1}$.  Including the next to leading
terms, the T-matrix has the structure
\begin{equation}
  \label{eq:T-low-kappa}
  T^{\textsc{mm}}_{lmlm}=\kappa^{2l}\bigg[\frac{(-1)^{l-1}(l+1)\alpha_l^\textsc{m}}{l (2l+1)!! (2l-1)!!}  \kappa 
+ \gamma^\textsc{m}_{l3}\kappa^{3}+\gamma^\textsc{m}_{l4}\kappa^{4} +\ldots\bigg] \, ,
\nonumber
\end{equation}
and $T^{\textsc{ee}}_{lmlm}$ is obtained by $\alpha^\textsc{m}_l \to
\alpha^\textsc{e}_l$,
$\gamma_{ln}^\textsc{m}\to\gamma^\textsc{e}_{ln}$.  The first terms
are
$\gamma^\textsc{m}_{13}=-[4+\mu(\epsilon\mu+\mu-6)]/[5(\mu+2)^2]R^5$,
$\gamma^\textsc{m}_{14}=(4/9)[(\mu-1)/(\mu+2)]^2R^6$, and
$\gamma^\textsc{e}_{13}$, $\gamma^\textsc{e}_{14}$ are obtained again
by interchanging $\mu$ and $\epsilon$. Higher order terms can be
easily obtained by expanding Eq.~(\ref{eq:t-matrix-elem-sphere}) for
small $\kappa$.

The translation matrix elements $\tilde \cU^{R,11}_{l'mlm}=
(-1)^{l'}\cU^{R,11}_{l'mlm}$ are obtained from
Eqs.~(\ref{eq:def-EM-U-matrix-mirror}),
(\ref{eq:EM-trans-mat-elements-z-B}) and
(\ref{eq:EM-trans-mat-elements-z-C}).  Using $h_l^{(1)}(iz)=-i^{-l}
\sqrt{2/(\pi z)} K_{l+1/2}(z)$ we get
\begin{widetext}
\begin{eqnarray}
  \label{eq:EM-trans-mat-elements-z-B-wK}
  B_{l'm'lm}(-2L\hat\bfz) &=& -\delta_{m'm} (-1)^{m} i^{l-l'} \sum_{l''} \frac{(-1)^{l''}}{2} 
\left[l(l+1)+l'(l'+1)-l''(l''+1)\right] \sqrt{\frac{(2l+1)(2l'+1)}{l(l+1)l'(l'+1)}}
(2l''+1)\nonumber\\
&&\times\begin{pmatrix}l&l'&l''\\0&0&0\end{pmatrix}
\begin{pmatrix}l&l'&l''\\m&-m&0\end{pmatrix} 
\frac{K_{l''+1/2}(2\kappa L)}{\sqrt{\pi \kappa L}}  \, ,\\
  \label{eq:EM-trans-mat-elements-z-C-wK}
C_{l'm'lm}(-2L\hat\bfz) &=& \delta_{m'm}  2 \kappa L m (-1)^m 
i^{l-l'} \sum_{l''} (-1)^{l''} 
\sqrt{\frac{(2l+1)(2l'+1)}{l(l+1)l'(l'+1)}}
(2l''+1)\nonumber\\
&&\times\begin{pmatrix}l&l'&l''\\0&0&0\end{pmatrix}
\begin{pmatrix}l&l'&l''\\m&-m&0\end{pmatrix}
\frac{K_{l''+1/2}(2\kappa L)}{\sqrt{\pi \kappa L}} \, .
\end{eqnarray}
\end{widetext}

Now we can employ the series representation of the Casimir
energy in Eq.~(\ref{eq:energy-scatt-exp}) with ${\mathbb N}=-\tilde
{\mathbb T}^1 \tilde {\mathbb U}^{R,11}$. Since the T-matrix is
diagonal in $l$, partial waves of order $l$ start to contribute to the
energy at order $L^{-2(l+1)}$. Also, the $p^\text{th}$ power of
${\mathbb N}$ becomes important only at order $L^{-(3p+1)}$. Notice
the stronger increase of the exponent with $p$ compared to the scalar
case where the exponent is $-(p+1)$. This can be understood from the
absence of $s$-waves for the electromagnetic field so that each
reflection contributes a factor $1/L^3$ due to $p$-waves. Matrix
operations and the expansion in $R/L$ are performed with {\tt
  Mathematica}.  From this we obtain for a dielectric sphere in front
of perfectly conducting mirror plane the Casimir  energy
\begin{widetext}
\begin{eqnarray}
  \label{eq:energy-eps-mu-sphere}
  \cE &=& -\frac{\hbar c}{\pi} \left\{
\frac{3}{8} (\alpha_1^\textsc{e} - \alpha_1^\textsc{m}) \frac{1}{L^4} 
+\frac{15}{32} (\alpha_2^\textsc{e} - \alpha_2^\textsc{m} +2 \gamma_{13}^\textsc{e} 
-2 \gamma_{13}^\textsc{m}) \frac{1}{L^6} 
+ \frac{1}{1024} \left[ 23  (\alpha_1^\textsc{m})^2 - 14  
\alpha_1^\textsc{m}  \alpha_1^\textsc{e}
+23  (\alpha_1^\textsc{e})^2 \right. \right. \nonumber\\
&+& \left. \left. 2160 (\gamma_{14}^\textsc{e}- \gamma_{14}^\textsc{m})
\right] \frac{1}{L^7}
+\frac{7}{7200} \left[ 572 (\alpha_3^\textsc{e}-\alpha_3^\textsc{m}) + 675 \left(
9(  \gamma_{15}^\textsc{e} -  \gamma_{15}^\textsc{m}) -55  
(  \gamma_{23}^\textsc{e} -  \gamma_{23}^\textsc{m})
\right)\right] \frac{1}{L^8} + \dots
\right\} \, .
\end{eqnarray}
\end{widetext}
The electric contribution to the leading term, $\sim L^{-4}$, was
obtained by Casimir and Polder for the interaction of an atom with
static polarizability $\alpha_1^\textsc{e}$ and a metallic surface
\cite{Casimir+48}. Later Boyer has generalized the leading order
result to include magnetic effects described by $\alpha_1^\textsc{m}$
\cite{Boyer69}. The higher order terms are new. They show how higher
order polarizabilities and frequency corrections to the static parameters
influence the interaction. There is no $\sim 1/L^5$ term. Notice also
that the first three terms of the contribution at order $L^{-7}$ have
precisely the structure of the Casimir-Polder interaction between two
atoms with static dipole polarizabilities $\alpha_1^\textsc{m}$ and
$\alpha_1^\textsc{e}$ but it is reduced by a factor of $1/2^8$. This
factor and the distance dependence $\sim L^{-7}$ of this term suggests
that it arises from the interaction of the dipole fluctuations inside
the sphere with those inside its image at a distance $2L$. The
additional coefficient of $1/2$ in the reduction factor $(1/2)(1/2^7)$
can be traced back to the fact that the forces involved in bringing
the dipole in from infinity act only on the dipole and not on its
image \cite{Barnett+00}.

For a perfectly conducting sphere the coefficients of the expansion of
the Casimir energy in $R/L$ become universal numbers. These
coefficients can be either obtained from
Eq.~(\ref{eq:energy-eps-mu-sphere}) by using the appropriate values of
the parameters for a prefect metal or it can be computed directly
from the T-matrix elements given in
Eqs.~(\ref{eq:t-matrix-elem-cond-sphere-m}),
(\ref{eq:t-matrix-elem-cond-sphere-e}). Following the latter route, 
we get the series
\begin{equation}
  \label{eq:EM-energy-series}
  \cE = \frac{\hbar c}{\pi} \frac{1}{L} \sum_{j=4}^\infty b_j \left(
\frac{R}{L}\right)^{j-1} \, ,
\end{equation}
where the coefficients up to order $1/L^{11}$ are
\begin{eqnarray}
\label{eq:coeff_EM}
b_4&=&-\frac{9}{16}, \quad
b_5=0, \quad b_6=-\frac{25}{32}, \quad b_7=-\frac{3023}{4096} 
\nonumber\\ 
\quad b_8&=&-\frac{12551}{9600},
\quad  b_9=\frac{1282293}{163840},\nonumber \\
b_{10}&=&-\frac{32027856257}{722534400},  
\,\,\, b_{11}=\frac{39492614653}{412876800} \, .
\end{eqnarray}
This and the corresponding results for a scalar field appear to be
asymptotic series. Therefore, the series cannot be summed to obtain
the interaction at small separations. In the next Section we use a
numerical implementation of Eq.~(\ref{eq:EM-casimir-energy-1-surface})
to compute the interaction at all separations.

\subsection{Non-perturbative result at all separations}

To obtain the Casimir interaction over a broad range of distances,
Eq.~(\ref{eq:EM-casimir-energy-1-surface}) has to be evaluated
numerically. As we employ a partial wave expansion, the computational
work increases with decreasing separation between the
objects. However, we shall see below that even at small separations
our method yields sufficient precision to obtain the leading
corrections to the proximity force approximation (PFA). The numerical
approach is based on the technique presented in
Refs.~\cite{Emig+07b,Emig+07}. Using the analytic expressions for the
matrix elements of the translation and transition matrices, we compute
the determinant and the integral over frequency $\kappa$ in
Eq.~(\ref{eq:EM-casimir-energy-1-surface}) numerically. The matrices
are truncated at a finite multipole order $l$ which yields a series of
estimates $\cE^{(l)}$ for the Casimir energy. The exact Casimir energy
is then obtained by extrapolating the series to $l\to\infty$. For the
geometry considered here, we observe an exponentially fast convergence
that allows us to obtain accurate results even at small separations
from a moderate multipole order. 

We compare our results for the interaction energy to the estimate that
follows from PFA. This is important since PFA is often used over a
range of separations although its accuracy is unknown even at
small distances for most geometries, including the electromagnetic
Casimir interaction between a plane and a sphere. The PFA estimate
for the latter geometry is given by
\begin{equation}
  \label{eq:PFA-result}
  \cE_\text{PFA} = \pi \Phi_0^{\pm/\text{EM}} \, \frac{\hbar c \, R}{d^2} 
\end{equation}
where $d=L-R$ is the surface-to-surface distance. The amplitudes follow
from the result for two parallel plates and are given by
\begin{eqnarray}
  \label{eq:PFA-amplitudes}
  \Phi_0^- &=& -\frac{\pi^2}{1440} \\
  \Phi_0^+ &=& +\frac{7 \pi^2}{11520} \\
  \Phi_0^\text{EM} &=& -\frac{\pi^2}{720}
\end{eqnarray}
for a scalar field with like ($-$) and unlike ($+$) boundary
conditions and the electromagnetic field, respectively.
In the following we present our results for the plane-sphere 
interaction over a wide range of separations for scalar fields
with Dirichlet and Neumann boundary conditions and for the
electromagnetic field.

\subsubsection{Scalar field}

\begin{figure}
\includegraphics[scale=0.9]{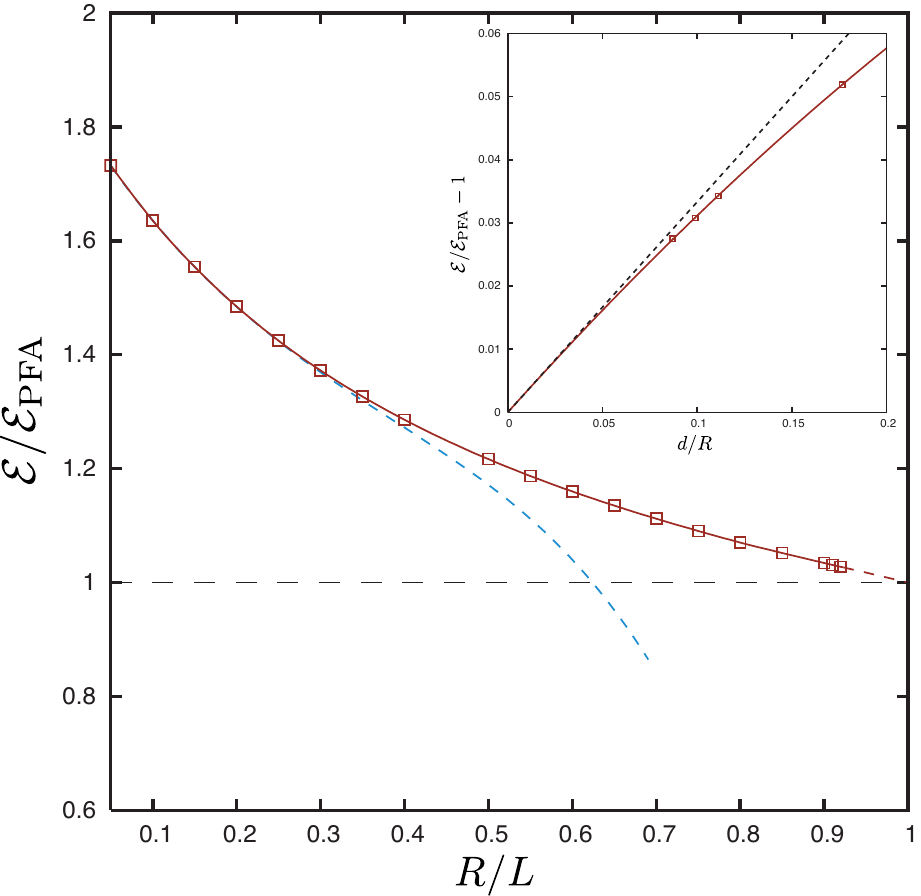}
\caption{Sphere-plane Casimir interaction of a scalar field with
  Dirichlet boundary conditions on both surfaces. The energy is scaled
  by the PFA estimate of Eq.~(\ref{eq:PFA-result}). The dashed curve
  represents the large distance expansion of
  Eq.~(\ref{eq:spheres-energy-large-L}) with the coefficients of
  Eq.~(\ref{eq:coeff_D-D}). Inset: Corrections to the PFA at small
  separations as function of the surface-to-surface distance
  $d=L-R$. The dashed curve corresponds to the lowest order correction
  to the PFA obtained by Bordag \cite{Bordag:qfext07}.}
\label{fig:DD}
\end{figure}
\begin{figure}
\includegraphics[scale=0.9]{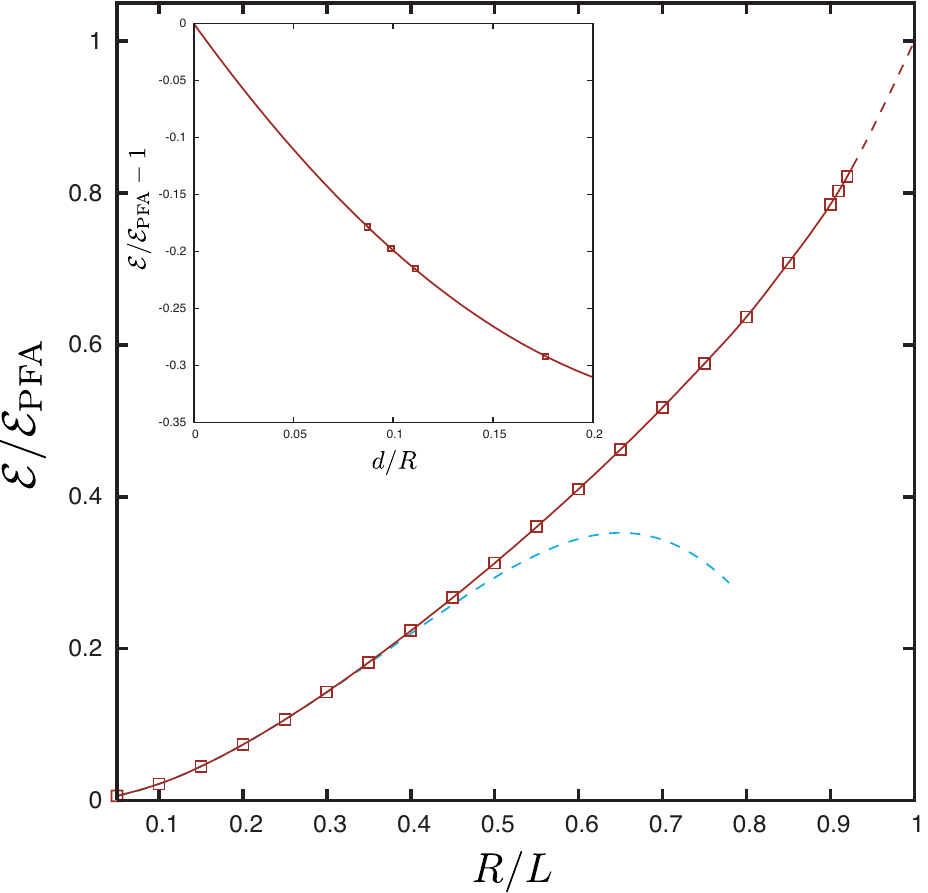}
\caption{Analog of Fig.~\ref{fig:DD} for Neumann boundary conditions
at the sphere and the plane. The dashed curve
  represents the large distance expansion of
  Eq.~(\ref{eq:spheres-energy-large-L}) with the coefficients of
  Eq.~(\ref{eq:coeff_N-N}).}
\label{fig:NN}
\end{figure}
\begin{figure}[!h]
\includegraphics[scale=0.9]{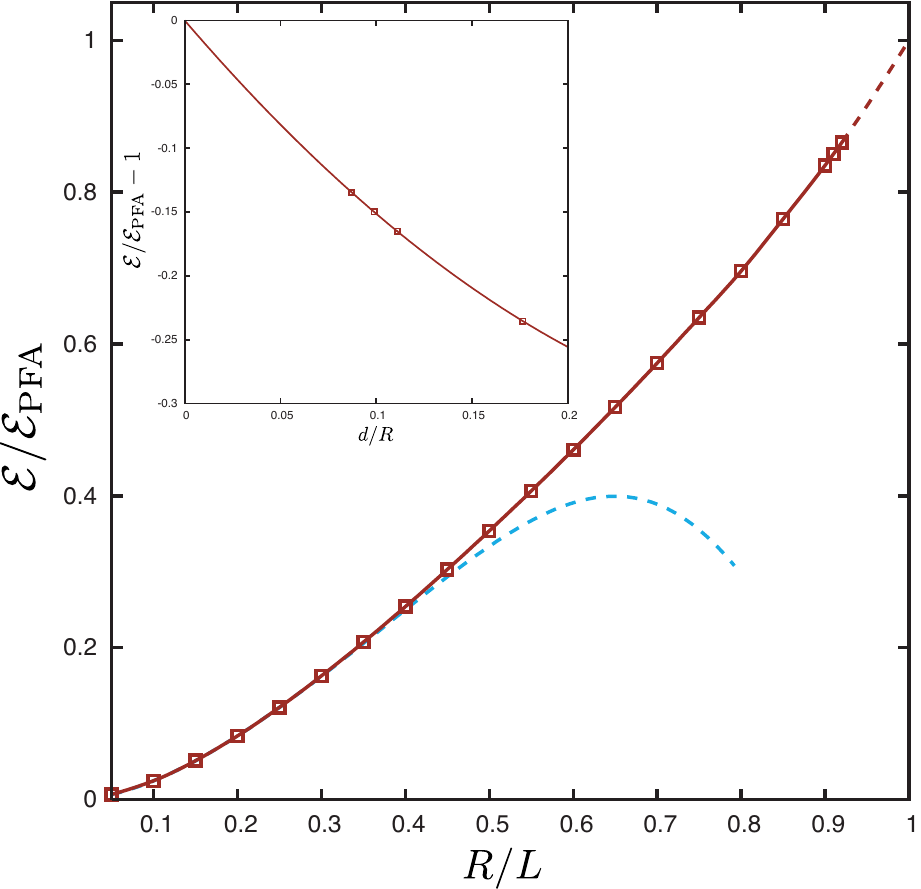}
\caption{Analog of Fig.~\ref{fig:DD} for Neumann boundary conditions
at the sphere and Dirichlet boundary conditions at the plane. The dashed curve
  represents the large distance expansion of
  Eq.~(\ref{eq:spheres-energy-large-L}) with the coefficients of
  Eq.~(\ref{eq:coeff_D-N}).}
\label{fig:ND}
\end{figure}
\begin{figure}[!h]
\includegraphics[scale=0.9]{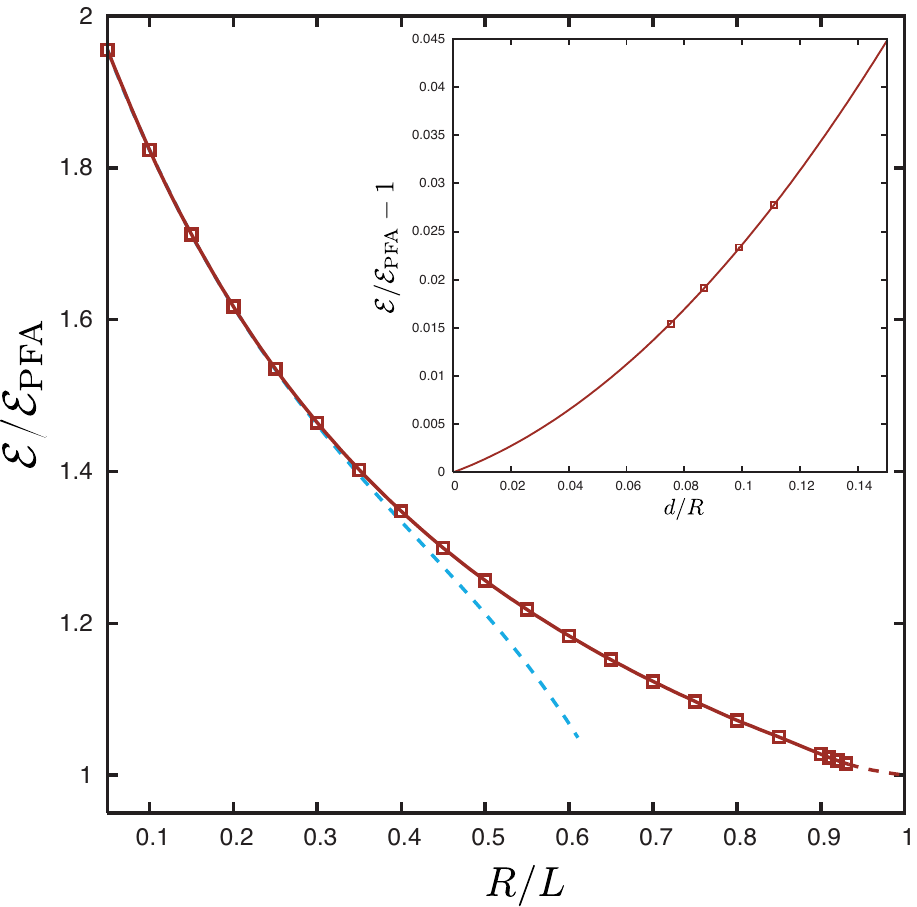}
\caption{Analog of Fig.~\ref{fig:DD} for Dirichlet boundary conditions
at the sphere and Neumann boundary conditions at the plane. The dashed curve
  represents the large distance expansion of
  Eq.~(\ref{eq:spheres-energy-large-L}) with the coefficients of
  Eq.~(\ref{eq:coeff_N-D}).}
\label{fig:DN}
\end{figure}

We consider the four cases that correspond to Dirichlet or Neumann
boundary conditions at the sphere and the mirror plane. The results
for the Casimir energy are shown in Figs.~\ref{fig:DD}-\ref{fig:DN}.
At large separations the results of the numerical evaluation of the
determinant agree nicely with the large distance expansions of the
previous section. For a sphere with Dirichlet boundary conditions, the
energy scales both at large and small separations as $R/d^2$ so that
the curves for $\cE/\cE_\text{PFA}$ in Figs.~\ref{fig:DD} and
\ref{fig:DN} tend also to a constant for $R/L\to 0$. The PFA
underestimates the actual interaction energy at all separations.  For
a Neumann sphere the interaction scales at large separations as
$R^3/L^4$ so that the curves for $\cE/\cE_\text{PFA}$ in
Figs.~\ref{fig:NN} and \ref{fig:ND} tend to zero for $R/L\to 0$. Here
the PFA overestimates the actual interaction energy at all
separations.

The non-perturbative approach allows us to study also the case of
small separations. In that limit our results can be fitted to a power
law of the form
\begin{equation}
  \label{eq:E-PFA+corrections}
  \cE = \cE_\text{PFA} \left[ 1+ \theta_1 \frac{d}{R} +  \theta_2 
\left(\frac{d}{R}\right)^2 + \ldots \right] \, .
\end{equation}
The coefficients $\theta_j$ obtained from a fit of the function of
Eq.~(\ref{eq:E-PFA+corrections}) to the data points for the four
smallest studied separations are summarized in Tab.~\ref{tab:thetas}.
The fitted curves are shown as insets in
Figs.~\ref{fig:DD}-\ref{fig:DN}.  The result for $\theta_1$ for the
case of Dirichlet boundary conditions at the sphere and the mirror agrees
with the analytical result $\theta_1=1/3$ presented in
Ref.~\cite{Bordag:qfext07}. Again for Dirichlet conditions, for the
second order coefficient a considerably larger numerical estimate of
$\theta_2=1.92$ has been obtained from world-line Monte Carlo sampling
\cite{Gies+06a,Gies+06b}. For the other combinations of boundary
conditions our findings represent the first results for the
corrections to the PFA.
\begin{table}
  \centering
  \begin{tabular}{|c|c|c|}
\hline
boundary condition & $\theta_1$ & $\theta_2$ \\
plane / sphere & & \\
\hline
 D / D & $0.3346 \pm 0.0017$ & $-0.231 \pm 0.012$ \\
 N / N & $-2.4298 \pm 0.0070$ & $4.394 \pm 0.049$ \\
 D / N & $-1.7493 \pm 0.0061$ & $2.354 \pm 0.042$ \\
 N / D & $0.1124 \pm 0.0043$ & $1.240 \pm 0.044$ \\
\hline
 EM / EM & $-1.42 \pm 0.02$ & $2.39 \pm 0.14$ \\
\hline    
  \end{tabular}
  \caption{Coefficients describing corrections to PFA, see 
Eq.~(\ref{eq:E-PFA+corrections}). D=Dirichlet, N=Neumann (scalar field) and 
EM=perfectly conducting boundary conditions (electromagnetic field).}
  \label{tab:thetas}
\end{table}

\subsubsection{Electromagnetic field}

\begin{figure}[!h]
\includegraphics[scale=0.9]{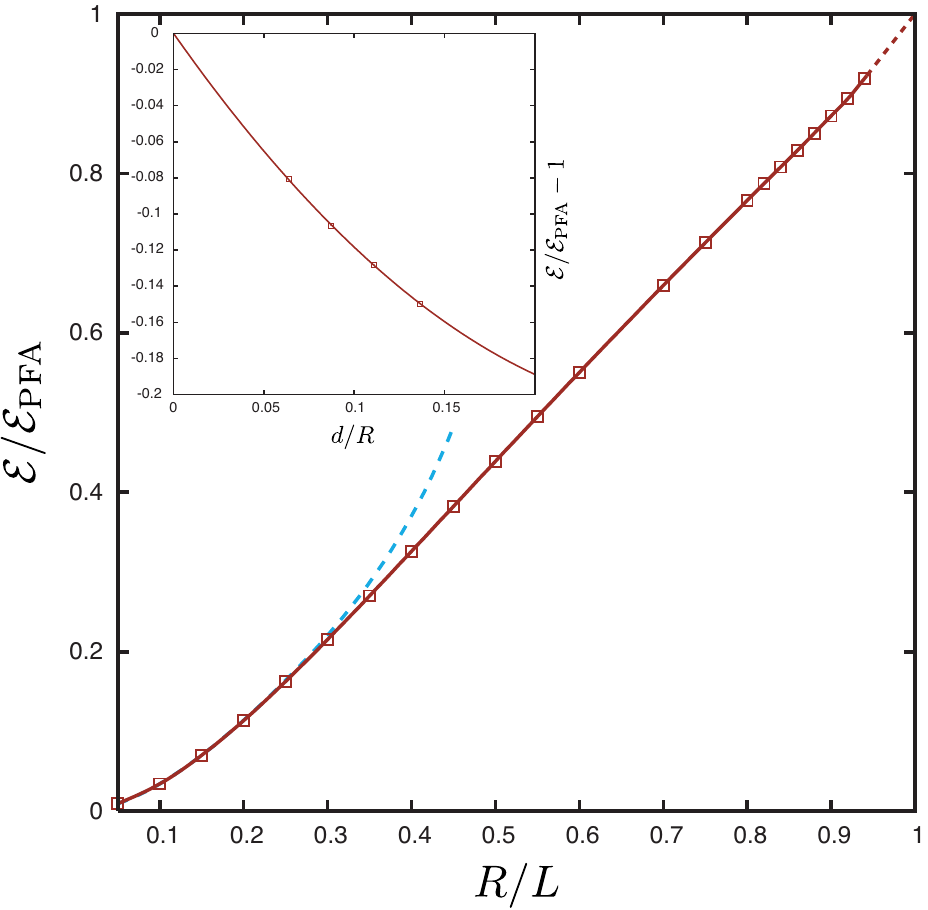}
\caption{Electromagnetic Casimir energy for the sphere-plate
  geometry. The energy is scaled by the PFA estimate of
  Eq.~(\ref{eq:PFA-result}). The asymptotic expansion of
  Eq.~(\ref{eq:EM-energy-series}) to order $1/L^{10}$ is shown as
  dashed line. Inset: Corrections to the PFA at small distances as
  function of $d=L-R$.}
\label{fig:EM}
\end{figure}

We focus on a perfectly conducting sphere and mirror plane.  The
Casimir energy resulting from a numerical computation of the
determinant of Eq.~(\ref{eq:EM-casimir-energy-1-surface}) is shown for
a wide range of separations in Fig.~\ref{fig:EM}. At large separations
the interaction is described by the asymptotic result of
Eq.~(\ref{eq:EM-energy-series}). At small distances the Casimir energy
approaches to PFA estimate and corrections to PFA can be described
again by Eq.~(\ref{eq:E-PFA+corrections}). A corresponding fit to the
data points for the four smallest separations is shown as inset in
Fig.~\ref{fig:EM}. The corresponding amplitudes of the correction terms
are listed in Tab.~\ref{tab:thetas}. 

This result is important for a number of recent measurements of the
Casimir force between almost flat surfaces. These experiments have
been performed for a plane mirror and a sphere with a radius that is
much larger than the distance between the surfaces in order to avoid
difficulties from parallelism control. Our results at short distances
indicate that for a sphere of radius $R=100\mu$m the corrections to
PFA are below $1\%$ only for surface-to-surface distances $d$ that are
smaller than $\approx 700$nm. For a sphere of that size, PFA fails
already by $10\%$ at a distance of $\approx 8\mu$m. From this we 
conclude that the currently achieved experimental accuracy for measurements
of Casimir forces in the sphere-plate geometry with a sphere of $R=100\mu$m
at distances below $d=1\mu$m is within the range or slightly less than 
the corrections to PFA. 

However, deviations from PFA become severe when smaller objects
interact with surfaces. Hence it is important to understand the
crossover of the Casimir interaction between macroscopic objects
and the eventual Casimir-Polder interaction between single atoms and a
surface. A description of this crossover is provided by our results in
Fig.~\ref{fig:EM}. Corresponding results can be also obtained for a
dielectric sphere or less symmetric objects by the methods presented
here. Finally, we note that it should be also possible to compute
correction amplitudes like the $\theta_j$ in
Eq.~\eqref{eq:E-PFA+corrections} analytically by applying methods
similar to those used in Ref. \cite{Bordag06}.  However, the validity
range of the corresponding series would be limited to small
separations and for an overall description of the interaction one
should resort to the device of a numerical evaluation along the lines
presented here.

\section{Acknowledgments} 
This work emerged from a collaboration with R.~L.~Jaffe, M.~Kardar,
and N.~Graham on related problems.  Support by the Heisenberg program
of the Deutsche Forschungsgemeinschaft is acknowledged.

\end{document}